\documentclass[preprint,authoryear,12pt]{elsarticle}
\usepackage{amssymb}

\usepackage{amsmath}
\usepackage{amsmath,color}
\usepackage{graphicx}
\usepackage{hyperref}

\newcommand{\ve}[1]{\mbox{\boldmath ${#1}$}}
\newcommand{\vesub}[2]{\mbox{{\boldmath ${#1}$}$_{#2}$}}

\begin{document}

\begin{frontmatter}
\title{Extended T-process Regression Models}
\author[a,b]{Zhanfeng Wang}
\address
[a]{Department of Statistics and Finance, University of Science and Technology of China, Hefei, China}
\author[c]{Jian Qing Shi}
\address
[c]{School of Mathematics and Statistics, Newcastle University, Newcastle, UK}
\author[b]{Youngjo Lee\corref{cor1}}
\cortext[cor1]{Corresponding author. Email: youngjo@snu.ac.kr}
\address[b]{Department of Statistics, Seoul National University, Seoul, Korea}
\begin{abstract}
Gaussian process regression (GPR) model has been widely used to fit data when the regression function is unknown  and its nice properties have been well established. In this article, we introduce an extended t-process regression (eTPR) model, which gives a robust best linear unbiased predictor (BLUP). Owing to its succinct construction, it inherits many attractive properties from the GPR model, such as having closed forms of marginal and predictive distributions to give an explicit form for robust BLUP procedures, and easy to cope with large dimensional covariates with an efficient implementation by slightly modifying existing BLUP procedures. Properties of the robust BLUP are studied. Simulation studies and real data applications show that the eTPR model gives a robust fit in the presence of outliers in both input and output spaces and has a good performance in prediction, compared with the GPR and locally weighted scatterplot smoothing (LOESS) methods.
\end{abstract}
\end{frontmatter}

\section{Introduction}

Consider a functional regression model
\begin{equation}
y_{i}=f_{0}(\mbox{\boldmath${x}$}_{i})+\epsilon _{i},~~i=1,...,n
\label{true}
\end{equation}
where $f_{0}(\mbox{\boldmath${x}$}_{i})$ is the value of unknown function $%
f_{0}(\cdot )$ at the $p\times 1$ observed covariate $%
\mbox{{\boldmath
${x}$}$_{i}$}\in \mathcal{X}=R^{p}$ and $\epsilon _{i}$ is an error term. To
fit an unknown function $f_{0},$ we may consider a process regression model
\begin{equation}
y(\mbox{\boldmath ${x}$})=f(\mbox{\boldmath${x}$})+\epsilon (%
\mbox{\boldmath
${x}$}),  \label{assumed}
\end{equation}
where $f(\mbox{\boldmath${x}$})$ is a random function and $\epsilon (%
\mbox{\boldmath
${x}$})$ is an error process for $\mbox{\boldmath ${x}$}\in \mathcal{X}$. A
GPR model assumes a Gaussian process (GP) for the random function $f(\cdot )$%
. It has been widely used to fit data when the regression function is
unknown: for detailed descriptions see Rasmussen and William (2006), and Shi
and Choi (2011) and references therein.
GPR has many good features, for example, it can model nonlinear relationship
nonparametrically between a response and a set of large dimensional
covariates with efficient implementation procedure. In this paper we
introduce an eTPR model and investigate advantages in using an extended
t-process (ETP).

BLUP procedures in linear mixed model are widely used (Robinson, 1991) and
extended to Poisson-gamma models (Lee and Nelder, 1996) and Tweedie models
(Ma and Jorgensen, 2007). Efficient BLUP algorithms have been developed for
genetics data (Zhou and Stephens, 2012) and spatial data (Dutta and Mondal,
2015). In this paper, we show that BLUP procedures can be extended to GPR
models. However, GPR fits are susceptible to outliers in output space ($%
y_{i} $). LOESS (Cleveland and Devlin, 1988) has been developed for a robust
fit against such outliers. However, it requires fairly large densely sampled
data set to produce good models and does not produce a regression function
that is easily represented by a mathematical formula. For models with many
covariates, it is inevitable to have sparsely sampled regions. Wauthier and
Jordan (2010) showed that the GPR model tends to give an overfit of data
points in the sparsely sampled regions (outliers in the input space, $%
\mbox{\boldmath${x}$}_{i}$). Thus, it is important to develop a method which
produces good fits for sparsely sampled regions as well as densely sampled regions. Wauthier and Jordan (2010)
proposed to use a heavy-tailed process. However, their copula method does
not lead to a close form for prediction of $f(\mbox{\boldmath${x}$}).$ As an
alternative to generate a heavy-tailed process, various forms of student $t$%
-process have been developed: see for example Yu \textit{et al}. (2007),
Zhang and Yeung (2010), Archambeau and Bach (2010) and Xu \textit{et al}.
(2011). However, Shah \textit{et al}. (2014) noted that the $t$-distribution
is not closed under addition to maintain nice properties in Gaussian models.

In this paper, we develop a specific eTPR model which is closed under addition to retain many favorable properties
of GPR models. Due to its special structure of
construction, the resulting eTPR model gives computationally efficient
algorithm, i.e. a slight modification of the
existing BLUP algorithm provides the robust BLUP procedure. Under the proposed eTPR model, marginal and predictive distributions are in closed
forms. Furthermore, it gives a robust BLUP procedure against outliers in
both input and output spaces. Properties of the robust BLUP procedure are investigated.

The remainder of the paper is as follows. Section 2 presents
an ETP and its properties. Section 3 proposes an eTPR model and discusses
the inference and implementation procedures. Robustness properties and
information consistency of robust BLUP predictions are shown in
Section 3. Numerical studies and real examples are in Section 4, followed by concluding remarks in Section 5. All the
proofs are in Appendix.

\section{Extended $t$-process}

\begin{figure}[h]
\begin{center}
\includegraphics[height = 0.6\textwidth,
width=0.95\textwidth]{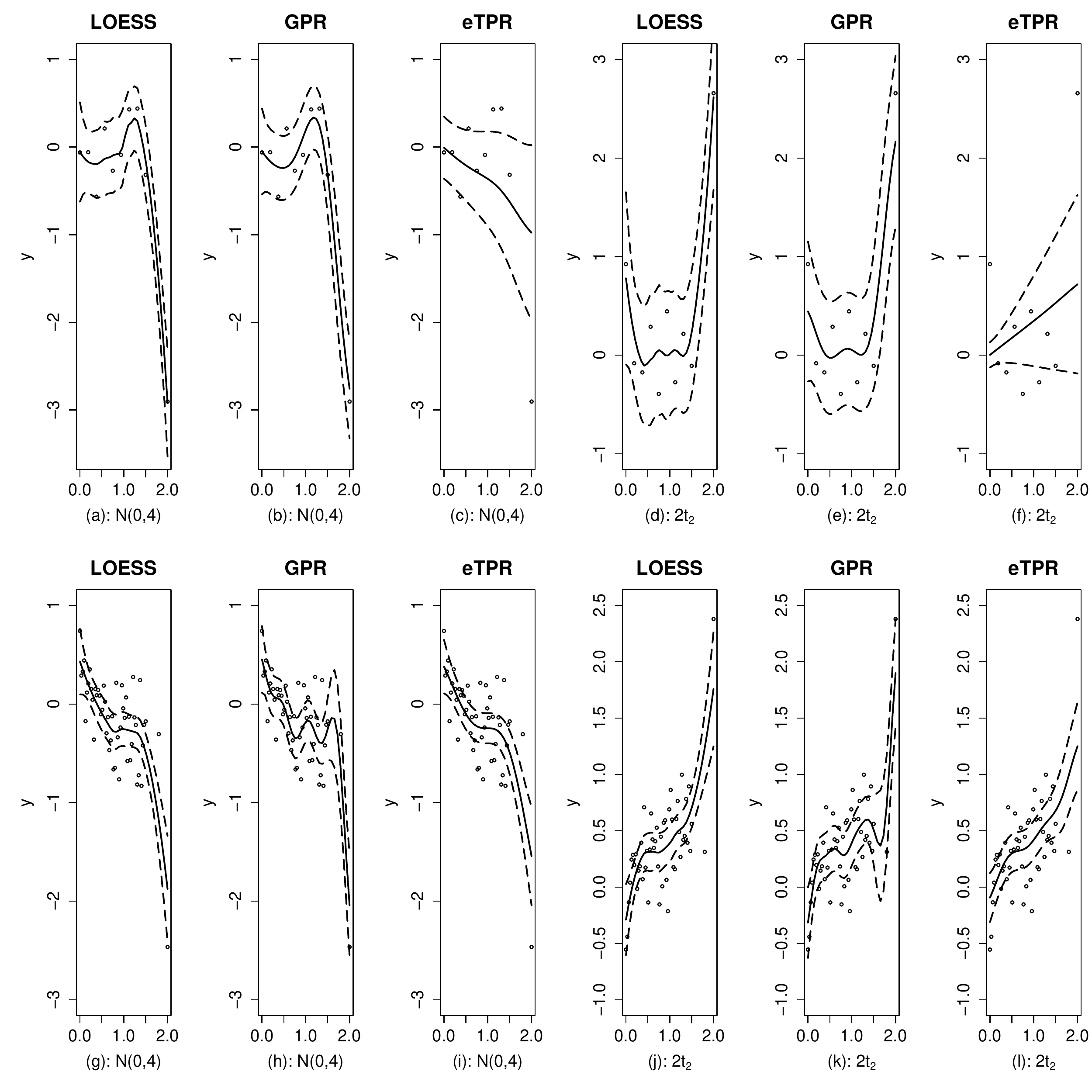}
\end{center}
\caption{Predictions in the presence of outliers at data point point 2.0
disturbed by additional errors with the normal distribution $N(0,4) $ or the
$t$ distribution $2 t_{2}$ where circles represent data points, solid and
dashed lines stand for predicted curves and their 95\% confidence bounds
from the LOESS, GPR and eTPR methods, respectively. }
\label{fig1}
\end{figure}

As a motivating example, we generated two data sets with sample sizes of $%
n=10$ and $n=50$ where $x_{i}$'s are evenly spaced in [0,~1.5] for the 9 (or
48) data points and the remaining point is at 2.0 (or two points at 1.8 and
2.0). Thus, the remaining point or the two points are sparse ones, meaning
they are far away from the other data points in input space. In addition, we also make the
data point 2.0 to be an outlier in output space by adding an extra error
from either $N(0,4)$ or $2t_{2}$, where $t_{2}$ is the student $t$
distribution with two degree of freedom. Prediction curves for simulated
data are plotted in Figure \ref{fig1}, where circles represent data points,
solid and dashed lines stand for prediction and their 95\% confidence
bounds. The true function is zero. For a small sample size $n=10$, Figure
1(a-f) shows that LOESS and GPR predictions are similar and the eTPR
prediction is the smoothest and shrinks the data point 2.0 the most heavily,
i.e. selective shrinkage occurs. For a moderate sample size $n=50$, Figure
1(g-l) shows that LOESS and eTPR predictions are similar. However, the eTPR
prediction still shrinks the most at 2.0. Even though unreported, for a
large sample size $n=100,$ all give similar predictions.

Denote observed data set by $\mbox{{\boldmath
${{\mathcal{D}}}$}$_{n}$}=\{\mbox{{\boldmath ${X}$}$_{n}$},%
\mbox{{\boldmath
${y}$}$_{n}$}\}$ where $\mbox{{\boldmath${y}$}$_{n}$}=(y_{1},...,y_{n})^{T}$
and $\mbox{{\boldmath${X}$}$_{n}$}=(\mbox
{{\boldmath${x}$}$_{1}$},...,\mbox{{\boldmath${x}$}$_{n}$})^{T}$. For random
component $f(\mbox{\boldmath ${u}$})$ at a new point $\mbox{\boldmath ${u}$}%
\in \mathcal{X}$, the best unbiased predictor is $E(f(\mbox{\boldmath
${u}$})|\mbox{{\boldmath
${{\mathcal{D}}}$}$_{n}$}).$ It is called a BLUP if it is linear in $%
\mbox{{\boldmath
${y}$}$_{n}$}.$ Its standard error can be estimated with $Var(f(%
\mbox{\boldmath ${u}$})|\mbox{{\boldmath
${{\mathcal{D}}}$}$_{n}$}).$ To have an efficient implementation procedure,
it is useful to have explicit forms for the predictive distribution $p(f(%
\mbox{\boldmath ${u}$})|\mbox{{\boldmath
${{\mathcal{D}}}$}$_{n}$}),$ $E(f(\mbox{\boldmath
${u}$})|\mbox{{\boldmath
${{\mathcal{D}}}$}$_{n}$})$ and $Var(f(\mbox{\boldmath ${u}$})|%
\mbox{{\boldmath
${{\mathcal{D}}}$}$_{n}$})$.

Let $f$ be a real-valued random function such that $f: {\mathcal{X}}\rightarrow R$.
Analogous to double hierarchical generalized linear models (Lee and Nelder,
2006), we consider a following hierarchical process,
\begin{equation*}
f|r\sim GP(h,rk),~~~~r\sim \mathrm{IG}(\nu ,\omega ),
\end{equation*}
where $GP(h,rk)$ stands for GP with mean function $h$ and covariance
function $rk$, and $\mathrm{IG}(\nu ,\omega )$ stands for an inverse gamma
distribution with the density function
\begin{equation*}
g(r)=\frac{1}{\Gamma (\nu )}(\frac{\omega }{r})^{\nu +1}\frac{1}{\omega }%
\exp {(-\frac{\omega }{r}),}
\end{equation*}
and $\Gamma (\cdot )$ is the gamma function. Then, $f$ follows an ETP $%
f\sim ETP(\nu ,\omega ,h,k),$ implying that for any collection of points $%
\mbox{\boldmath
${X}$}_{n}=(\mbox{{\boldmath${x}$}$_{1}$},...,\mbox{{\boldmath${x}$}$_{n}$}%
)^{T},\mbox{{\boldmath${x}$}$_{i}$}\in {\mathcal{X}}$, we have
\begin{equation*}
\mbox{{\boldmath${f}$}$_{n}$}=f(\mbox{{\boldmath ${X}$}$_{n}$})=(f(%
\mbox{{\boldmath${x}$}$_{1}$}),...,f(\mbox {{\boldmath${x}$}$_{n}$}%
))^{T}\sim EMTD(\nu ,\omega ,\mbox{{\boldmath
${h}$}$_{n}$},\mbox{{\boldmath${K}$}$_{n}$}),
\end{equation*}
where $\mbox{{\boldmath ${f}$}$_{n}$}\sim EMTD(\nu ,\omega ,%
\mbox{{\boldmath
${h}$}$_{n}$},\mbox{{\boldmath${K}$}$_{n}$})$ means that $%
\mbox{{\boldmath
${f}$}$_{n}$}$ has an extended multivariate $t$-distribution (EMTD) with the
density function,
\begin{equation*}
p(z)=|2\pi \omega \mbox{{\boldmath ${K}$}$_{n}$}|^{-1/2}\frac{\Gamma
(n/2+\nu )}{\Gamma (\nu )}\left( 1+\frac{(z-\mbox{{\boldmath ${h}$}$_{n}$}%
)^{T}\mbox{{\boldmath ${K}$}$_{n}^{-1}$}(z-\mbox{{\boldmath ${h}$}$_{n}$})}{%
2\omega }\right) ^{-(n/2+\nu )},
\end{equation*}
$\mbox{{\boldmath${h}$}$_{n}$}=(h(\mbox{{\boldmath${x}$}$_{1}$}),...,h(%
\mbox{{\boldmath${x}$}$_{n}$}))^{T}$, $\mbox{{\boldmath
${K}$}$_{n}$}=(k_{ij})_{n\times n}$ and $k_{ij}=k(%
\mbox{{\boldmath
${x}$}$_{i}$},\mbox{{\boldmath${x}$}$_{j}$})$ for some mean function $%
h(\cdot ): {\mathcal{X}}\rightarrow R$ and kernel function $k(\cdot ,\cdot ):  {\mathcal{X}}\times
 {\mathcal{X}}\rightarrow R.$

It follows that at any collection of finite points ETP has an analytically
representable EMTD density being similar to GP having multivariate normal
density. Note that $E(\mbox{{\boldmath${f}$}$_{n}$})=%
\mbox
{{\boldmath${h}$}$_{n}$}$ is defined when $\nu >1/2$ and $Cov(%
\mbox
{{\boldmath${f}$}$_{n}$})=\omega \mbox{{\boldmath${K}$}$_{n}$}/(\nu -1)$ is
defined when $\nu >1$. When $\nu =\omega =\alpha /2$, $%
\mbox{{\boldmath
${f}$}$_{n}$}$ becomes the multivariate $t$-distribution of Lange \textit{et
al}. (1989). When $\nu =\alpha /2$ and $\omega =\beta /2$, $%
\mbox{{\boldmath
${f}$}$_{n}$}$ becomes the generalized multivariate $t$-distribution of
Arellano-Valle and Bolfarine (1995). For $f\sim ETP(\nu ,\omega ,0,k)$
it easily obtains that $E(f(\mbox{\boldmath${x}$}))=0,~~Var(f(%
\mbox{\boldmath${x}$}))={\omega }k(\mbox{\boldmath${x}$},\mbox{%
\boldmath${x}$})/{(\nu -1)}$, and
\begin{align}
& Skewness(f(\mbox{\boldmath${x}$}))=\frac{E(f^{3}(\mbox{\boldmath${x}$}))}{%
(E(f^{2}(\mbox{\boldmath${x}$})))^{3/2}}=0,  \notag \\
& Kurtosis(f(\mbox{\boldmath${x}$}))=\frac{E(f^{4}(\mbox{\boldmath${x}$}))}{%
(E(f^{2}(\mbox{\boldmath${x}$})))^{2}}=\frac{3}{\nu -2}+3\geq 3\text{ when }%
\nu >2.  \notag
\end{align}
Thus, we may say that the $ETP(\nu ,\omega ,0,k)$ has a heavier tail than
the $GP(0,k)$.

\vskip10pt \noindent \textbf{Proposition 1} \textit{Let $f\sim ETP(\nu
,\omega ,h,k).$ }

\begin{itemize}
\item[(i)]  \textit{When $\omega /\nu \rightarrow \lambda $ as $\nu
\rightarrow \infty $, we have $\lim_{\nu \rightarrow \infty }ETP(\nu ,\omega
,h,k)=GP(h,\lambda k).$}

\item[(ii)]  \textit{Let $\mbox{\boldmath${Z}$}\in \mathcal{X}$ be a p$%
\times 1$ random vector such that $\mbox{\boldmath${Z}$}\sim EMTD(\nu
,\omega ,\mbox{{\boldmath${\mu}$}$_{z}$},\mbox{{\boldmath${\Sigma}$}$_{z}$})$%
. For a linear system $f(\mbox{\boldmath ${x}$})=%
\mbox{{\boldmath
${x}$}$^{T}$}\mbox{\boldmath ${Z}$}$ with $\mbox{\boldmath ${x }$}\in
\mathcal{X}$, we have $f\sim ETP(\nu ,\omega ,h,k)$ with $h(%
\mbox{\boldmath
${x}$})=\mbox{{\boldmath ${x}$}$^{T}$}\mbox{{\boldmath
${\mu}$}$_{z}$}$ and $k(\mbox{{\boldmath ${x}$}$_{i}$},%
\mbox{{\boldmath
${x}$}$_{j}$})=\mbox{{\boldmath ${x}$}$_{i}^{T}$}%
\mbox{{\boldmath
${\Sigma}$}$_{z}$}\mbox{{\boldmath ${x}$}$_{j}$}$ }.

\item[(iii)]  \textit{\ Let $\mbox{\boldmath ${u}$}\in \mathcal{X}$ be a new
data point and $\mbox{{\boldmath ${k}$}$_{\ve u}$}=(k(\mbox{\boldmath${u}$},%
\mbox{{\boldmath${x}$}$_{1}$}),...,k(\mbox{\boldmath${u}$},%
\mbox{{\boldmath${x}$}$_{n}$}))^{T}.$ Then, $f|%
\mbox{{\boldmath
${f}$}$_{n}$}\sim ETP(\nu ^{\ast },\omega ^{\ast },h^{\ast },k^{\ast })$
with $\nu ^{\ast }=\nu +n/2,$ $\omega ^{\ast }=\omega +n/2$,
\begin{align}
& h^{\ast }(\mbox{\boldmath${u}$})=\mbox {{\boldmath${k}$}$_{\ve u}^{T}$}%
\mbox{{\boldmath${K}$}$_{n}^{{-1}}$}(\mbox{{\boldmath${f}$}$_{n}$}-%
\mbox{{\boldmath${h}$}$_{n}$})+h(\mbox {\boldmath${u}$}),  \notag \\
& \mathit{k^{\ast }(\mbox{\boldmath ${u}$},\mbox{\boldmath ${v}$})}=\frac{{%
2\omega +(\mbox {{\boldmath${f}$}$_{n}$}-\mbox{{\boldmath${h}$}$_{n}$})^{T}%
\mbox {{\boldmath${K}$}$_{n}^{{-1}}$}(\mbox{{\boldmath${f}$}$_{n}$}-\mbox
{{\boldmath${h}$}$_{n}$})}}{{2\omega +n}}\ \left( k(\mbox{\boldmath
${u}$},\mbox{\boldmath ${v}$})-\mbox{{\boldmath
${k}$}$_{\ve u}^{T}$}\mbox{{\boldmath${K}$}$_{n}^{{-1}}$}\mbox{{%
\boldmath${k}$}$_{\ve v}$}\right) ,  \notag
\end{align}
for $\mbox{\boldmath ${v }$}\in \mathcal{X}$.}
\end{itemize}

Even if the mean and covariance functions of $f\sim ETP(\nu ,\omega ,h,k)$
cannot be defined when $\nu <0.5$, from Proposition 1(iii), the mean and
covariance functions of the conditional process $f|%
\mbox{{\boldmath
${f}$}$_{n}$}$ do always exist if $n\geq 2$. Also from Proposition 1(iii),
the conditional process $ETP(\nu ^{\ast },\omega ^{\ast },h^{\ast },k^{\ast
})$ converges to a GP, as either $\nu $ or $n$ goes to $\infty $. Thus, if
the sample size $n$ is large enough, the ETP behaves like a GP.

For a new point $\mbox{\boldmath ${u}$}$, we have $f(\mbox{\boldmath ${u}$})|%
\mbox{{\boldmath ${f}$}$_{n}$}\sim EMTD(\nu ^{\ast },\omega ^{\ast },h^{\ast
}(\mbox{\boldmath ${u}$}),k^{\ast }(\mbox{\boldmath ${u}$},%
\mbox{\boldmath
${u}$}))$, where
\begin{align}
& h^{\ast }(\mbox{\boldmath ${u}$})=E(f(\mbox{\boldmath ${u}$})|%
\mbox{{\boldmath ${f}$}$_{n}$})=\mbox {{\boldmath${k}$}$_{\ve u}^{T}$}%
\mbox{{\boldmath
${K}$}$_{n}^{{-1}}$}(\mbox{{\boldmath${f}$}$_{n}$}-%
\mbox{{\boldmath
${h}$}$_{n}$})+h(\mbox{\boldmath ${u}$}),  \notag \\
& Var(f(\mbox{\boldmath ${u}$})|\mbox{{\boldmath ${f}$}$_{n}$})=\frac{\omega
^{\ast }}{\nu ^{\ast }-1}k^{\ast }(\mbox{\boldmath ${u}$},%
\mbox{\boldmath
${u}$})=s~\{k(\mbox{\boldmath ${u}$},\mbox{\boldmath ${u}$})-%
\mbox{{\boldmath
${k}$}$_{\ve u}^{T}$}\mbox{{\boldmath${K}$}$_{n}^{{-1}}$}\mbox{{%
\boldmath${k}$}$_{\ve u}$}\},  \notag
\end{align}
and $s=({2\omega +(\mbox{{\boldmath${f}$}$_{n}$}-\mbox{{%
\boldmath${h}$}$_{n}$})^{T}\mbox{{\boldmath${K}$}$_{n}^{{-1}}$}(%
\mbox{{\boldmath${f}$}$_{n}$}-\mbox{{\boldmath${h}$}$_{n}$})})/({2\nu +n-2})$%
. Note that from Lemma 2(iv) $s=E(r|\mbox{{\boldmath${f}$}$_{n}$})$.

Under various combinations of $\nu $ and $\omega $, the ETP generates
various $t$-processes proposed in the literature. For example, $ETP(\alpha
/2,\alpha /2-1,h,k)$ is the $t$-process of Shah \textit{et al}. (2014). They
showed that if covariance function $\mbox{\boldmath ${\Sigma }$}$ follows an
inverse Wishart process with parameter $\alpha =2\nu $ and kernel function $%
k $, and $f|\mbox{\boldmath ${\Sigma }$}\sim GP(h,(\alpha -2)%
\mbox{\boldmath
${\Sigma }$})$, then $f$ has an extended $t$-process $ETP(\alpha /2,\alpha
/2-1,h,k)$. $ETP(\alpha /2,\alpha /2,h,k)$ is the Student's $t$-process of
Rasmussen and William (2006) and $ETP(\nu ,1/2,h,k)$ is that of Zhang and
Yeung (2010).

\section{eTPR models}

Consider the process regression model (\ref{assumed})
\begin{equation*}
y(\mbox{\boldmath ${x}$})=f(\mbox{{\boldmath ${x}$}})+\epsilon (%
\mbox{\boldmath ${x}$}),~~for~~\mbox{\boldmath ${x}$}\in \mathcal{X}.
\end{equation*}
In this
section we introduce an eTPR model, where $f$ and $\epsilon $ have a joint
ETP process,
\begin{equation}
\left(
\begin{array}{c}
f \\
\epsilon
\end{array}
\right) \sim ETP\left( \nu ,\omega ,\left(
\begin{array}{c}
h \\
0
\end{array}
\right) ,\left(
\begin{array}{cc}
k & 0 \\
0 & \tilde{k}
\end{array}
\right) \right) ,  \label{assum1}
\end{equation}
kernel function $\tilde{k}(\mbox{\boldmath ${u}$},\mbox{\boldmath ${v}$}%
)=\phi I(\mbox{\boldmath ${u}$}=\mbox{\boldmath ${v}$})$ and $I(\cdot )$ is
an indicator function. The joint ETP above can be constructed
hierarchically as
\begin{equation*}
\left(
\begin{array}{c}
f \\
\epsilon
\end{array}
\right) \Big|r\sim GP\left( \left(
\begin{array}{c}
h \\
0
\end{array}
\right) ,r\left(
\begin{array}{cc}
k & 0 \\
0 & \tilde{k}
\end{array}
\right) \right) {\mbox{~ and~ }}r\sim \mathrm{{IG}(\nu ,\omega ),}
\end{equation*}
and this implies that $f+\epsilon |r\sim GP(h,r(k+\tilde{k}))$ and $r\sim
\mathrm{{IG}(\nu ,\omega )}$
to give $y\sim ETP(\nu ,\omega ,h,k+\tilde{k}).$
Hence, additivity property of the GP and many other properties hold
conditionally and marginally in the ETP. When $r=1$, the eTPR model becomes a GPR model.

For observed data, this leads to a functional regression model
\begin{equation*}
y_{i}=f(\mbox{{\boldmath ${x}$}$_{i}$})+\epsilon _{i},~~i=1,...,n,
\end{equation*}
where $y_{i}=y(\mbox{{\boldmath ${x}$}$_{i}$})$ and $\epsilon _{i}=\epsilon (%
\mbox{{\boldmath ${x}$}$_{i}$})$. Now it follows that
\begin{align}
& f(\mbox{{\boldmath ${X}$}$_{n}$})|\mbox{{\boldmath ${X}$}$_{n}$}\sim
EMTD(\nu ,\omega ,\mbox{{\boldmath
${h}$}$_{n}$},\mbox{{\boldmath ${K}$}$_{n}$}),  \notag \\
& \mbox{{\boldmath
${y}$}$_{n}$}|f,\mbox{{\boldmath ${X}$}$_{n}$}\sim EMTD(\nu ,\omega ,f(%
\mbox{{\boldmath ${X}$}$_{n}$}),\phi \mbox{{\boldmath ${I}$}$_{n}$}),  \notag
\\
& \mbox{{\boldmath ${y}$}$_{n}$}|\mbox{{\boldmath ${X}$}$_{n}$}\sim EMTD(\nu
,\omega ,\mbox{{\boldmath ${h}$}$_{n}$},%
\mbox{{\boldmath
${\tilde\Sigma}$}$_{n}$}),  \notag
\end{align}
where $\mbox{{\boldmath ${\tilde\Sigma}$}$_{n}$}=%
\mbox{{\boldmath
${K}$}$_{n}$}+\phi \mbox{{\boldmath ${I}$}$_{n}$}$.

Consider a linear mixed model
\begin{equation*}
y_{i}=\mbox{{\boldmath ${w}$}$_{i}^{T}$}\mbox{\boldmath ${\delta }$}+%
\mbox{{\boldmath ${v}$}$_{i}^{T}$}\mbox{\boldmath ${b}$}+\epsilon
_{i},~~i=1,...,n,
\end{equation*}
where \mbox{{\boldmath ${w}$}$_{i}$} is the design matrix for fixed effects $%
\mbox{\boldmath ${\delta}$}$, $\mbox{{\boldmath ${v}$}$_{i}$}$ is the design
matrix for random effect $\mbox{\boldmath ${b}$}\thicksim N(0,\theta %
\mbox{{\boldmath ${I}$}$_{p}$})$ and $\epsilon _{i}\thicksim N(0,\phi )$ is
a white noise. Suppose that $\mbox{{\boldmath ${X}$}$_{n}$}=(%
\mbox{{\boldmath
${W}$}$_{n}$},\mbox{{\boldmath ${V}$}$_{n}$})$, $\ f(%
\mbox{{\boldmath
${X}$}$_{n}$})=\mbox{{\boldmath ${W}$}$_{n}^{T}$}\mbox{\boldmath ${\delta}$}+%
\mbox{{\boldmath ${V}$}$_{n}^{T}$}\mbox{\boldmath ${b
}$}$, $\mbox{{\boldmath ${h}$}$_{n}$}=\mbox{{\boldmath ${W}$}$_{n}^{T}$}%
\mbox{\boldmath ${\delta}$}$ and $\mbox{{\boldmath ${K}$}$_{n}$}=\theta %
\mbox{{\boldmath ${V}$}$_{n}$}\mbox{{\boldmath ${V}$}$_{n}^{T}$}$ with $%
\mbox{{\boldmath ${W}$}$_{n}$}=(\mbox{{\boldmath ${w}$}$_{1}$},...,%
\mbox{{\boldmath ${w}$}$_{n}$})^{T}$ and $\mbox{{\boldmath ${V}$}$_{n}$}=(%
\mbox{{\boldmath ${v}$}$_{1}$},...,\mbox{{\boldmath ${v}$}$_{n}$})^{T}$.
Then, the linear mixed model becomes the functional regression model with
\begin{align}
& f(\mbox{{\boldmath ${X}$}$_{n}$})|\mbox{{\boldmath ${X}$}$_{n}$}=%
\mbox{{\boldmath ${W}$}$_{n}^{T}$}\mbox{\boldmath ${\delta}$}+%
\mbox{{\boldmath ${V}$}$_{n}^{T}$}\mbox{\boldmath ${b}$}|%
\mbox{{\boldmath
${X}$}$_{n}$}\sim N(\mbox{{\boldmath ${W}$}$_{n}^{T}$}%
\mbox{\boldmath ${\delta
}$},\mbox{{\boldmath ${K}$}$_{n}$}),  \notag \\
& \mbox{{\boldmath
${y}$}$_{n}$}|f,\mbox{{\boldmath ${X}$}$_{n}$}=%
\mbox{{\boldmath
${y}$}$_{n}$}|\mbox{\boldmath ${b}$},\mbox{{\boldmath ${X}$}$_{n}$}\sim N(%
\mbox{{\boldmath ${W}$}$_{n}^{T}$}\mbox{\boldmath ${\delta}$}+%
\mbox{{\boldmath ${V}$}$_{n}^{T}$}\mbox{\boldmath ${b
}$},\phi \mbox{{\boldmath ${I}$}$_{n}$}),  \notag \\
& \mbox{{\boldmath ${y}$}$_{n}$}|\mbox{{\boldmath ${X}$}$_{n}$}\sim N(%
\mbox{{\boldmath ${W}$}$_{n}^{T}$}\mbox{\boldmath ${\delta }$},%
\mbox{{\boldmath
${\tilde\Sigma}$}$_{n}$}).  \notag
\end{align}
This shows that eTPR models extend the conventional normal linear mixed
models to a nonlinear functional regression. In contrary to LOESS, this also
shows that the eTPR method can produce a regression function, easily
represented by a mathematical formula.

In the hierarchical construction of ETP, there is only one single random
effect $r,$ so that $r$ is not estimable, confounded with parameters in
covariance matrix. This means that $\nu $ and $\omega $ are not estimable.
Following Lee and Nelder (2006), we set $\omega =\nu -1$ because $%
Var(f)=\omega k/(\nu -1)=k$ if $f\sim ETP(\nu ,\omega ,h,k).$ Thus, the
variance does not depend upon $\nu $ as $Var(f)=k;$ this is also true when $%
f\sim GP(h,k).$ By doing this way, the first two moments of GP and ETP have
the same parametrization. Zellner (1976) also noted that $\nu $ cannot be
estimated with a single realization of $\{(y_{i},\mbox{{%
\boldmath${x}$}$_{i}$}):i=1,2,...,n\}$. In multivariate t-distribution,
Lange \textit{et al}. (1989) proposed to use $\nu =2$. Zellner (1976)
suggested that $\nu $ can be chosen according to investigator's knowledge of
robustness of regression error distribution. As $\nu \rightarrow \infty $,
ETP tends to GP. When robustness property is an important issue, a smaller $%
\nu $ is preferred. We tried various values for $v$ and find that $v=1.05$
works well. From now on we set $v=1.05$ to have $\omega =\nu -1=0.05>0.$
Furthermore, in functional regression models it is conventional to assume $h(%
\mbox{\boldmath ${u}$})=0.$ Thus, without loss of generality we assume $h(%
\mbox{\boldmath ${u}$})=0.$

\subsection{Parameter estimation for eTPR}

{So far we have assumed that the covariance kernel $k(\cdot ,\cdot )$ is
given. To fit the eTPR model, we need to choose $k(\cdot ,\cdot )$. A way is
to estimate the covariance kernel nonparametrically; see e.g. Hall \textit{%
et al}. (2008). However, this method is very difficult to be applied to
problems with multivariate covariates. Thus, we choose a covariance kernel
from a function family such as a squared exponential kernel and Mat\'{e}rn
class kernel.} This paper employs a combination of a squared exponential and
a non-stationary linear covariance kernel as follows,
\begin{equation}
k(\mbox{{\boldmath ${x}$}$_{i}$},\mbox{{\boldmath ${x}$}$_{j}$};%
\mbox{\boldmath ${\theta}$})=\theta _{0}\exp {\left( -\frac{1}{2}%
\sum_{l=1}^{p}\theta _{1l}(x_{i,l}-x_{j,l})^{2}\right) }+\sum_{l=1}^{p}%
\theta _{2l}x_{i,l}x_{j,l},  \label{kerfun}
\end{equation}
where $\mbox{\boldmath ${\theta}$}=\{\theta _{0},\theta _{1l},\theta
_{2l},l=1,...,p\}$ are a set of hyper-parameters. In (\ref{kerfun}), $%
1/\theta _{1l}$ measure the length scale of each input covariate, $\theta
_{0}$ known as scaling parameter which controls the vertical scale of
variations of a typical function of the input, and $\theta _{2l}$ defines
the scale of non-stationary linear trends. The small value of $1/\theta
_{1l} $ means that the corresponding covariate may have great contribution
in the covariance function. More about kernel function $k(\cdot ,\cdot ;%
\mbox{\boldmath
${\theta}$})$ can be seen in Rasmussen and William (2006) and Shi and Choi
(2011).

Let $\mbox{\boldmath ${\beta}$}=(\phi ,\mbox{\boldmath ${\theta}$})$ where $%
\phi $ is a parameter for $\epsilon (\mbox{\boldmath ${x}$})$ and $%
\mbox{\boldmath ${\theta}$}$ are those for $f(\mbox{{\boldmath ${x}$}})$.
Here the joint density of $\mbox{{\boldmath ${y}$}$_{n}$},f(%
\mbox{{\boldmath
${X}$}$_{n}$})|\mbox{{\boldmath ${X}$}$_{n}$}$ is
\begin{equation}
p_{\beta }(\mbox{{\boldmath ${y}$}$_{n}$},f(\mbox{{\boldmath ${X}$}$_{n}$})|%
\mbox{{\boldmath ${X}$}$_{n}$})=p_{\phi }(\mbox{{\boldmath ${y}$}$_{n}$}|f,%
\mbox{{\boldmath ${X}$}$_{n}$})p_{\theta }(f(\mbox{{\boldmath ${X}$}$_{n}$})|%
\mbox{{\boldmath
${X}$}$_{n}$}),  \notag
\end{equation}
where $p_{\phi }(\mbox{{\boldmath ${y}$}$_{n}$}|f,%
\mbox{{\boldmath
${X}$}$_{n}$})$ and $p_{\theta }(f(\mbox{{\boldmath
${X}$}$_{n}$})|\mbox{{\boldmath
${X}$}$_{n}$})$ are density functions of EMTDs. Because $%
\mbox{{\boldmath
${y}$}$_{n}$}|\mbox{{\boldmath ${X}$}$_{n}$}\sim EMTD(\nu ,\nu -1,0,%
\mbox{{\boldmath
${\tilde\Sigma}$}$_{n}$})$, the maximum likelihood (ML) estimator $\hat{%
\mbox{\boldmath ${\beta}$}}$ for $\mbox{\boldmath ${\beta}$}$ can be
obtained by solving
\begin{equation}
\frac{\partial \log p{_{\beta }(\mbox{{\boldmath
${y}$}$_{n}$}|\mbox{{\boldmath ${X}$}$_{n}$})}}{\partial
\mbox{\boldmath
${\beta}$}}=\frac{1}{2}Tr\left( \Big(s_{1}\mbox{\boldmath
${\alpha}$}\mbox{{\boldmath
${\alpha}$}$^{T}$}-\mbox{{\boldmath ${\tilde{\Sigma}}$}$_{n}^{-1}$}\Big)%
\frac{\partial \mbox{{\boldmath ${\tilde{\Sigma}}$}$_{n}$}}{\partial %
\mbox{\boldmath ${\beta}$}}\right) =0,  \notag
\end{equation}
where $\mbox{\boldmath ${\alpha}$}=%
\mbox{{\boldmath
${\tilde{\Sigma}}$}$_{n}^{-1}$}\mbox{{\boldmath ${y}$}$_{n}$}$, $s_{1}=({n+2\nu })/({2(\nu -1)+\mbox{{\boldmath ${y}$}$_{n}^{T}$}%
\mbox{{\boldmath
${\tilde{\Sigma}}$}$_{n}^{-1}$}\mbox{{\boldmath ${y}$}$_{n}$}})$, and
\begin{equation}
p_{\beta }(\mbox{{\boldmath ${y}$}$_{n}$}|\mbox{{\boldmath ${X}$}$_{n}$}%
)=|2\pi (\nu -1)\mbox{{\boldmath
${\tilde{\Sigma}}$}$_{n}$}|^{-1/2}\frac{\Gamma (n/2+\nu )}{\Gamma (\nu )}%
\left( 1+\frac{\mbox{{\boldmath ${y}$}$_{n}^{T}$}%
\mbox{{\boldmath
${\tilde\Sigma}$}$_{n}^{-1}$}\mbox{{\boldmath
${y}$}$_{n}$}}{2(\nu -1)}\right) ^{-(n/2+\nu )}.  \label{margin}
\end{equation}
 The score equations for GPR models above are ML estimating
equations in linear mixed models with $\nu =\infty $ and $s_{1}=1.$ Thus, a
little modification of existing BLUP procedures gives a parameter estimation
for eTPR models.

\subsection{Predictive distribution}

Since
\begin{equation}
\left(
\begin{array}{c}
f(\mbox{{\boldmath ${X}$}$_{n}$}) \\
\mbox{{\boldmath ${y}$}$_{n}$}
\end{array}
\right) \Bigg|\mbox{{\boldmath ${X}$}$_{n}$}\sim EMTD\left( \nu ,\nu
-1,0,\left(
\begin{array}{cc}
\mbox{{\boldmath ${K}$}$_{n}$} & \mbox{{\boldmath ${K}$}$_{n}$} \\
\mbox{{\boldmath ${K}$}$_{n}$} & \mbox{{\boldmath ${\tilde\Sigma}$}$_{n}$}
\end{array}
\right) \right) ,  \notag
\end{equation}
from Lemma 2(iii) we have $f(\mbox{{\boldmath
${X}$}$_{n}$})|\mbox{{\boldmath
${\cal D}$}$_{n}$}=\{\vesub{X}{n},\vesub{y}{n}\}\thicksim EMTD(n/2+\nu ,n/2+\nu -1,%
\mbox
{{\boldmath${\mu}$}$_{n}$},\mbox{{\boldmath${\Sigma}$}$_{n}$}),$ with
\begin{align}
& \mbox{{\boldmath${\mu}$}$_{n}$}=E(f(\mbox{{\boldmath ${X}$}$_{n}$})|%
\mbox{{\boldmath ${\cal D}$}$_{n}$})=\mbox{{\boldmath${K}$}$_{n}$}%
\mbox{{\boldmath ${\tilde\Sigma}$}$_{n}^{-1}$}\mbox{{\boldmath${y}$}$_{n}$},
\notag \\
& \mbox{{\boldmath${\Sigma}$}$_{n}$}=Cov(f(\mbox{{\boldmath ${X}$}$_{n}$})|%
\mbox{{\boldmath ${\cal D}$}$_{n}$})=s_{0}\phi \mbox{{\boldmath${K}$}$_{n}$}%
\mbox{{\boldmath ${\tilde\Sigma}$}$_{n}^{-1}$},  \notag \\
& s_{0}=E(r|\mbox{{\boldmath ${\cal D}$}$_{n}$})=\frac{%
\mbox{{\boldmath
${y}$}$_{n}^{T}$}\mbox{{\boldmath
${\tilde\Sigma}$}$_{n}^{-1}$}\mbox{{\boldmath
${y}$}$_{n}$}+2(\nu -1)}{n+2(\nu -1)}.  \notag
\end{align}
Thus, given $\mbox{\boldmath ${\beta }$},$ $E(f(%
\mbox{{\boldmath
${X}$}$_{n}$})|\mbox{{\boldmath ${\cal D}$}$_{n}$})$ is linear in $%
\mbox{{\boldmath${y}$}$_{n}$},$ i.e. the BLUP for $f(%
\mbox{{\boldmath
${X}$}$_{n}$}),$ which is an extension of the BLUP in linear mixed models to eTPR models. This BLUP
has a form independent of $v,$ so that it is the BLUP for GPR
models. However, the conditional variance depends upon $v,$ except when $%
r=1, $ i.e. $s_{0}=1$ under GPR models. Thus, the BLUPs for the eTPR and GPR
models have a common form, but have different predictors and their variance
estimations because of different parameter estimations ($s_{0}\neq 1$ and $%
s_{1}\neq 1$). Furthermore, all quantities necessary to compute $s_{0}$ and $%
s_{1}$\ are available during implementing BLUP procedures.

For a given new data point $\mbox{\boldmath ${u}$}$, we have
\begin{equation*}
\left(
\begin{array}{c}
\mbox{{\boldmath${y}$}$_{n}$} \\
f(\mbox{\boldmath ${u}$})
\end{array}
\right)\Bigg|\mbox{{\boldmath ${X}$}$_{n}$} \sim EMTD\left( \nu ,\nu
-1,0,\left(
\begin{array}{cc}
\mbox{{\boldmath ${\tilde\Sigma}$}$_{n}$} &
\mbox{{\boldmath ${k}$}$_{\ve
u}$} \\
\mbox{{\boldmath ${k}$}$_{\ve u}^{T}$} & k(\mbox{\boldmath ${u}$},%
\mbox{\boldmath ${u}$})
\end{array}
\right) \right) .
\end{equation*}
By Lemma 2(iii), the predictive distribution $p(f( \mbox{\boldmath
${u}$})|\mbox{{\boldmath
${\mathcal{D}}$}$_{n}$})$ is $EMTD(n/2+\nu ,n/2+\nu -1,\mu _{n}^{\ast
},\sigma _{n}^{\ast }),$ where
\begin{align}
& \mu _{n}^{\ast }=E(f(\mbox{\boldmath ${u}$})|%
\mbox{{\boldmath
${\mathcal{D}}$}$_{n}$})=\mbox{{\boldmath ${k}$}$_{\ve u}^{T}$}%
\mbox{{\boldmath
${\tilde{\Sigma}}$}$_{n}^{{-1}}$}\mbox{{\boldmath${ y}$}$_{n}$},
\label{spec.mean} \\
& \sigma _{n}^{\ast }=Var(f(\mbox{\boldmath ${u}$})|%
\mbox{{\boldmath
${\mathcal{D}}$}$_{n}$})=s_{0}\Big(k(\mbox{\boldmath ${u}$},%
\mbox{\boldmath ${u
}$})-\mbox{{\boldmath ${k}$}$_{\ve u}^{T}$}\mbox{{\boldmath${\tilde{%
\Sigma}}$}$_{n}^{-1}$}\mbox{{\boldmath ${k}$}$_{\ve u}$}\Big).
\label{spec.cov}
\end{align}
Furthermore, from Proposition 1(iii), ${\ f|%
\mbox{{\boldmath ${
\mathcal{D}}$}$_{n}$}}\sim ETP{(n/2+\nu ,n/2+\nu -1,\ h^{\ast },k^{\ast }),\
} $ {where $h^{\ast }(\mbox{\boldmath ${u}$})=%
\mbox{{\boldmath
${\mu}$}$_{n}^{*}$}$ and $k^{\ast }(\mbox{\boldmath${u}$},%
\mbox{\boldmath${v}$})=s_{0}\Big(k(\mbox
{\boldmath${u}$},\mbox{\boldmath${v}$})-k_{\mbox{\boldmath${u}$}}^{T}%
\mbox{{\boldmath${\tilde{\Sigma}}$}$_{n}^{-1}$}k_{\mbox{\boldmath${v}$}}\Big)%
. $ } From Lemma 2(iii), we also have
$
y(\mbox{\boldmath ${u}$})|\mbox{{\boldmath${\mathcal{D}}$}$_{n}$}\thicksim
EMTD(n/2+\nu ,n/2+\nu -1,\mu _{n}^{\ast },\sigma _{n}^{\ast }+s_{0}\phi )
$
with $E(y(\mbox{\boldmath ${u}$})|\mbox{{\boldmath${\mathcal{D}}$}$_{n}$}%
)=\mu _{n}^{\ast }$ and $Var(y(\mbox{\boldmath ${u}$})|\mbox{{\boldmath${%
\mathcal{D}}$}$_{n}$})=\sigma _{n}^{\ast }+s_{0}\phi $. Consequently, this
conditional predictive process can be used to construct prediction {$\hat{y}(%
\mbox{\boldmath ${u}$})=E(y(\mbox{\boldmath ${u}$})|%
\mbox{{\boldmath
${\mathcal{D}}$}$_{n}$})=$}$E(f(\mbox{\boldmath ${u}$})|%
\mbox{{\boldmath
${\mathcal{D}}$}$_{n}$})${{\ of the unobserved response $y(%
\mbox{\boldmath
${u}$})$ at }}$\mbox{\boldmath ${x}$}=\mbox{\boldmath
${u}$}${\ and its standard error can be formed using the predictive
variance, given by} $\sigma _{n}^{\ast }+s_{0}\phi $, and the proof is in
Appendix. The predictive variance for $\hat{f}(\mbox{\boldmath ${u}$})$
in (\ref{spec.cov}) differs from that for $\hat{y}(\mbox{\boldmath ${u}$}).$

The prediction of $f(\mbox{{\boldmath ${X}$}$_{n}$})$ and $f(%
\mbox{\boldmath
${u}$})$ discussed above is the best unbiased predictions under eTPR models,
and so is under GPR models. However, their standard errors (variance
estimators) differ. Note that
\begin{equation*}
s_{0}=\frac{\mbox{{\boldmath
${y}$}$_{n}^{T}$}\mbox{{\boldmath
${\tilde\Sigma}$}$_{n}^{-1}$}\mbox{{\boldmath
${y}$}$_{n}$}+2(\nu -1)}{n+2(\nu -1)}=\frac{(\mbox{{\boldmath ${y}$}$_{n}$}-%
\hat{\mbox{\boldmath ${f}$}}_{n})^{T}\mbox{{\boldmath${\tilde\Sigma}$}$_{n}$}%
(\mbox{{\boldmath ${y}$}$_{n}$}-\hat{\mbox{\boldmath ${f}$}}_{n})/\phi
^{2}+2(\nu -1)}{n+2(\nu -1)},
\end{equation*}
where $\hat{\mbox{\boldmath ${f}$}}_{n}$ is the BLUP\ for $f(%
\mbox{{\boldmath ${X}$}$_{n}$})$. Thus, the standard error estimate of the
BLUP under the eTPR model increases if the model does not fit the responses $%
\mbox{{\boldmath ${y}$}$_{n}$}$ well while that under the GPR model does not
depend upon the model fit.

Random-effect models consist with three objects, namely the data $%
\mbox{{\boldmath
${\mathcal{D}}$}$_{n}$}$, unobservables (random effects) and parameters
(fixed unknowns) $\mbox{\boldmath ${\beta}$}.$\ For inferences of such
models, Lee and Nelder (1996) proposed the use of the h-likelihood.\ Lee and
Kim (2015) showed that inferences about unobservables allow both Bayesian
and frequentist interpretations. In this paper, we see that the eTPR model
is an extension of random-effect models. Thus, we may view the functional
regression model (\ref{assumed}) either as a Bayesian model, where a GP or
an ETP as a prior, or as a frequentist model where a latent process such as
GP and ETP is used to fit unknown function $f_{0}()$ in a functional space
(Chapter 9, Lee \textit{et al}., 2006). With the predictive distribution
above, we may form both Bayesian credible and frequentist confidence
intervals. Estimation procedures in Section 3.1\ can be viewed as an
empirical Bayesian method with a uniform prior on $%
\mbox{\boldmath
${\beta}$}.$ In frequentist (or Bayesian) approach, (\ref{margin}) is a
marginal likelihood for fixed (or hyper) parameters.

\subsection{Robust properties}

Let $\hat{f}_{T}(\mbox{\boldmath ${u}$})=\hat{\mbox{\boldmath ${\mu}$}}%
_{n}^{\ast }=\mbox{{\boldmath ${\mu}$}$_{n}^{*}$}|_{%
\mbox{\boldmath
${\beta}$}=\hat{\mbox{\boldmath ${\beta}$}}}$ and $V_{T}=\hat{%
\mbox{\boldmath
${\sigma}$}}_{n}^{\ast }=\mbox{{\boldmath ${\sigma}$}$_{n}^{*}$}|_{%
\mbox{\boldmath ${\beta}$}=\hat{\mbox{\boldmath ${\beta}$}}}$ be the BLUP
for $f(\mbox{\boldmath ${u}$})$ and its variance estimate, respectively,
under the eTPR model. And let $\hat{f}_{G}(\mbox{\boldmath ${u}$})$ and $%
V_{G} $ be those under the GPR model with $s_{0}=1$. Let $M_{T}=(\hat{f}_{T}(%
\mbox{\boldmath ${u}$})-f_{0}(\mbox{\boldmath ${u}$}))/\sqrt{V_{T}}$ and $%
M_{G}=(\hat{f}_{G}(\mbox{\boldmath ${u}$})-f_{0}(\mbox{\boldmath ${u}$}))/%
\sqrt{V_{G}}$ be two student t-type statistics for a null hypothesis $f(%
\mbox{\boldmath ${u}$})=f_{0}(\mbox{\boldmath ${u}$})$. Under a bounded
kernel function, if $y_{i}\rightarrow \infty $ for some $i$, $%
M_{G}\rightarrow \infty $, while $M_{T}$ remains bounded. Therefore, $M_{T}$
for eTPR is more robust against outliers in output space compared to that
for GPR. This property still holds for ML estimators.

\vskip10pt \noindent \textbf{Proposition 2} \textit{If kernel function $k(\mbox{\boldmath ${u}$},%
\mbox{\boldmath
${v}$};\mbox{\boldmath
${\theta}$})$ is bounded, continuous and differentiable on $%
\mbox{\boldmath${\theta}$}$, then the ML estimator $\hat{%
\mbox{\boldmath
${\beta}$}}$ from the eTPR has bound influence function, while that from the
GPR does not. }

\subsection{Information Consistency}

Let $p_{\phi _{0}}(\mbox{{\boldmath ${y}$}$_{n}$}|f_{0},%
\mbox{{\boldmath
${X}$}$_{n}$})$ be the density function to generate the data $%
\mbox{{\boldmath ${y}$}$_{n}$}$ given $\mbox{{\boldmath
${X}$}$_{n}$}$ under the true model (\ref{true}), where $f_{0}$ is the true
underlying function of $f$. Let $p_{\theta }(f)$ be a measure of random
process $f$ on space ${\mathcal{F}}=\{f(\cdot):\mathcal{X}\rightarrow R\}$.\
Let
\begin{equation*}
p_{\phi ,\theta }(\mbox{{\boldmath ${y}$}$_{n}$}|%
\mbox{{\boldmath
${X}$}$_{n}$})=\int_{{\mathcal{F}}}p_{\phi }(\mbox{{\boldmath
${y}$}$_{n}$}|f,\mbox{{\boldmath
${X}$}$_{n}$})dp_{\theta }(f),
\end{equation*}
be the density function to generate the data $\mbox{{\boldmath ${y}$}$_{n}$}$
given $\mbox{{\boldmath
${X}$}$_{n}$}$ under the assumed eTPR model (\ref{assum1}). Thus, the
assumed model (\ref{assum1}) is not the same as the true underlying model (%
\ref{true}). Here $\phi $ is the common in both models and $\phi_0$ is the true value of $\phi$. Let $p_{\phi
_{0},\hat \theta}(\mbox{{\boldmath
${y}$}$_{n}$}|\mbox{{\boldmath ${X}$}$_{n}$})$ be the estimated density
function under the eTPR model. Denote $D[p_{1},p_{2}]=\int (\log {p_{1}}%
-\log {p_{2}})dp_{1}$ by the Kullback-Leibler distance between two densities
$p_{1} $ and $p_{2}$. Then, we
have the following proposition.

\vskip10pt \noindent \textbf{Proposition 3} \textit{Under the appropriate conditions in Appendix, we have
\begin{equation*}
\frac{1}{n}E_{\mbox{{\boldmath ${X}$}$_{n}$}}(D[p_{\phi _{0}}(%
\mbox{{\boldmath
${y}$}$_{n}$}|f_{0},\mbox{{\boldmath ${X}$}$_{n}$}),p_{\phi _{0},\hat{\theta}%
}(\mbox{{\boldmath ${y}$}$_{n}$}|\mbox{{\boldmath ${X}$}$_{n}$}%
)])\longrightarrow 0,{\mbox{as}}~~n\rightarrow \infty ,
\end{equation*}
where the expectation is taken over the distribution of $%
\mbox{{\boldmath
${X}$}$_{n}$}$. }

\vskip 10pt

From Proposition 3, the Kullback-Leibler distance between two density
functions for $\mbox{{\boldmath ${y}$}$_{n}$}|\mbox{{\boldmath
${X}$}$_{n}$}$ from the true and the assumed models becomes zero,
asymptotically. Let $\mbox{{\boldmath ${y}$}$_{i}$}=(y_{1},...,y_{i})^{T}$
and $\mbox{{\boldmath ${X}$}$_{i}$}=(\mbox{{\boldmath ${x}$}$_{1}$},...,%
\mbox{{\boldmath ${x}$}$_{i}$})^{T}$, $i=1,...,n$. In Appendix, we show that
\begin{equation}\label{bspred}
p_{\phi _{0},\theta }(\mbox{{\boldmath ${y}$}$_{n}$}|%
\mbox{{\boldmath
${X}$}$_{n}$})=\prod_{i=1}^{n}p_{\phi _{0},\theta }(y_{i}|%
\mbox{{\boldmath
${X}$}$_{i}$},\mbox{{\boldmath ${y}$}$_{i-1}$}),
\end{equation}
where
\begin{align}
& p_{\phi _{0},\theta }(y_{i}|\mbox{{\boldmath
${X}$}$_{i}$},\mbox{{\boldmath ${y}$}$_{i-1}$})=\int_{{\mathcal{F}}}p_{\phi
_{0}}(y_{i}|f,\mbox{{\boldmath ${X}$}$_{i}$},\mbox{{\boldmath ${y}$}$_{i-1}$}%
)dp_{\theta }(f|\mbox{{\boldmath ${X}$}$_{i}$},%
\mbox{{\boldmath
${y}$}$_{i-1}$}),  \notag \\
& p_{\theta }(f|\mbox{{\boldmath ${X}$}$_{i}$},%
\mbox{{\boldmath
${y}$}$_{i-1}$})=\frac{p_{\phi _{0}}(\mbox{{\boldmath ${y}$}$_{i-1}$}|f,%
\mbox{{\boldmath
${X}$}$_{i-1}$})}{\int_{{\mathcal{F}}}p_{\phi _{0}}(%
\mbox{{\boldmath
${y}$}$_{i-1}$}|f^{\prime },\mbox{{\boldmath ${X}$}$_{i-1}$})dp_{\theta
}(f^{\prime })}.  \notag
\end{align}
Under the true model (\ref{true}), similarly to (\ref{bspred}), we have
\begin{equation*}
p_{\phi _{0}}(\mbox{{\boldmath ${y}$}$_{n}$}|f_{0},%
\mbox{{\boldmath
${X}$}$_{n}$})=\prod_{i=1}^{n}p_{\phi _{0}}(y_{i}|f_{0},%
\mbox{{\boldmath
${X}$}$_{i}$},\mbox{{\boldmath ${y}$}$_{i-1}$}).
\end{equation*}
Seeger \textit{et al}. (2008) called $p_{\phi _{0}}(y_{i}|f_{0},%
\mbox{{\boldmath
${X}$}$_{i}$},\mbox{{\boldmath ${y}$}$_{i-1}$})$ and $p_{\phi _{0},\hat{%
\theta}}(y_{i}|\mbox{{\boldmath
${X}$}$_{i}$},\mbox{{\boldmath ${y}$}$_{i-1}$})$ Bayesian prediction
strategies. We can show that
\begin{equation*}
D[p_{\phi _{0}}(\mbox{{\boldmath
${y}$}$_{n}$}|f_{0},\mbox{{\boldmath ${X}$}$_{n}$}),p_{\phi _{0},\hat{\theta}%
}(\mbox{{\boldmath ${y}$}$_{n}$}|\mbox{{\boldmath ${X}$}$_{n}$})]=\int
\sum_{i=1}^{n}Q(y_{i}|\mbox{{\boldmath
${X}$}$_{i}$},\mbox{{\boldmath ${y}$}$_{i-1}$})p_{\phi _{0}}(%
\mbox{{\boldmath ${y}$}$_{n}$}|f_{0},\mbox{{\boldmath ${X}$}$_{n}$})d%
\mbox{{\boldmath ${y}$}$_{n}$},
\end{equation*}
where $Q(y_{i}|\mbox{{\boldmath
${X}$}$_{i}$},\mbox{{\boldmath ${y}$}$_{i-1}$})=\log \{p_{\phi
_{0}}(y_{i}|f_{0},\mbox{{\boldmath
${X}$}$_{i}$},\mbox{{\boldmath ${y}$}$_{i-1}$})/p_{\phi _{0},\hat{\theta}%
}(y_{i}|\mbox{{\boldmath
${X}$}$_{i}$},\mbox{{\boldmath ${y}$}$_{i-1}$})\}$ is a loss function and $%
\sum_{i=1}^{n}Q(y_{i}|\mbox{{\boldmath
${X}$}$_{i}$},\mbox{{\boldmath ${y}$}$_{i-1}$})$ is called cumulative loss.
Under the GPR model, Seeger \textit{et al}. (2008)\ and Wang and Shi (2014)
proved Proposition 3, interpreted it as the average of cumulative loss $%
\sum_{i=1}^{n}Q(y_{i}|\mbox{{\boldmath
${X}$}$_{i}$},\mbox{{\boldmath ${y}$}$_{i-1}$})/n$ tending to zero
asymptotically, and called it the information consistency. In this paper, we
show this property for the robust BLUPs. Consequently, the frequentist BLUP
procedure is consistent with the Bayesian strategy in terms of average risk
over an ETP prior.

\section{Numerical studies}

\subsection{Simulation studies}

We use simulation studies to evaluate performance of the BLUP procedures
from the eTPR model (\ref{assum1}). For GPR and eTPR models, we use

\begin{itemize}
\item  GPR: $f\sim GP(0,k)$ and $\epsilon _{i}\sim N(0,\phi )$;

\item  eTPR: $f\sim ETP(\nu ,\nu -1,0,k)$ and $\epsilon \sim ETP(\nu ,\nu
-1,0,\tilde{k})$ ;
\end{itemize}

\noindent where kernel function $k$ is given in (\ref{kerfun}) and $\tilde{k}%
(u,v)=\phi I(u=v)$. Results are based on 500 simulation data.

\noindent \textbf{Selective shrinkage}

When some sparse data points are far away from the dense data points,
predictions of the sparse ones from the eTPR method are more heavily
regularized than those from the LOESS and the GPR methods. To generate data,
from the process model (\ref{assumed}) we assume $f$ follows a GP with mean $%
0$ and the kernel function (\ref{kerfun}), and error term follows a normal
distribution with mean 0 and variance $\phi $, denoted by $N(0,\phi )$. We
set $\mbox{\boldmath
${\beta}$}=(\phi ,\theta _{0},\theta _{11},\theta
_{21})=(0.1,~0.05,~10,~0.05)$. In Figure \ref{fig1}, 95\%
prediction confidence bounds are computed as $\hat{f}(u)\pm 1.96\sqrt{Var(%
\hat{f}(u))}$. At sparse data point, from Figure 1 we see that the eTPR
method has selective shrinkage of Wauthier and Jordan (2010) and gives a
wider interval.


We compare prediction performance from the LOESS, GPR and eTPR methods in
Table \ref{tab1}, where $n=10$ and the data point 2.0 is added with an extra error from
either $N(0,\sigma ^{2})$ or $\sigma t_{2}$ with $\sigma ^{2}=1$, 2, 3 and
4. Testing data points are evenly spaced from interval (0,~2.0), denoted by $%
\{x_{j}^{\ast }:j=1,...,m\}$ with $m=30$. Prediction performance of the test
data points is measured with mean squared error $MSE=\sum_{j=1}^{m}\hat{f}%
(x_{j}^{\ast })^{2}/m$. Table \ref{tab1} shows that robust BLUPs from the
eTPR model have the smallest MSE among the three methods: LOESS, GPR and
eTPR. The improvement is greater with $t$ error.

\begin{table}[!ht]
\caption{Mean squared errors of prediction results and their standard
deviation (in parentheses) by the LOESS, GPR and eTPR methods with $
\mbox{\boldmath
${\beta}$}=(0.1,~0.05,~0.05,~10)$.}
\label{tab1}\tabcolsep=5pt \fontsize{10}{16}\selectfont
 \vskip 0pt
\par
\begin{center}
\begin{tabular}{ccccc}
error & $\sigma^2$ & LOESS & GPR & eTPR \\ \hline
Normal & 1 & 0.238(0.236) & 0.204(0.255) & 0.167(0.220) \\
& 2 & 0.330(0.363) & 0.305(0.397) & 0.235(0.333) \\
& 3 & 0.421(0.493) & 0.406(0.533) & 0.300(0.440) \\
& 4 & 0.513(0.623) & 0.507(0.669) & 0.357(0.539) \\ \cline{2-5}
t & 1 & 0.735(3.183) & 0.721(3.142) & 0.390(1.298) \\
& 2 & 1.054(3.063) & 0.974(2.106) & 0.518(1.086) \\
& 3 & 1.498(4.575) & 1.372(3.053) & 0.818(3.462) \\
& 4 & 1.646(3.269) & 1.614(3.165) & 0.859(1.998) \\ \hline
\end{tabular}
\end{center}
\end{table}

\begin{figure}[h!]
\begin{center}
\includegraphics[height=0.6\textwidth,width=0.95\textwidth]{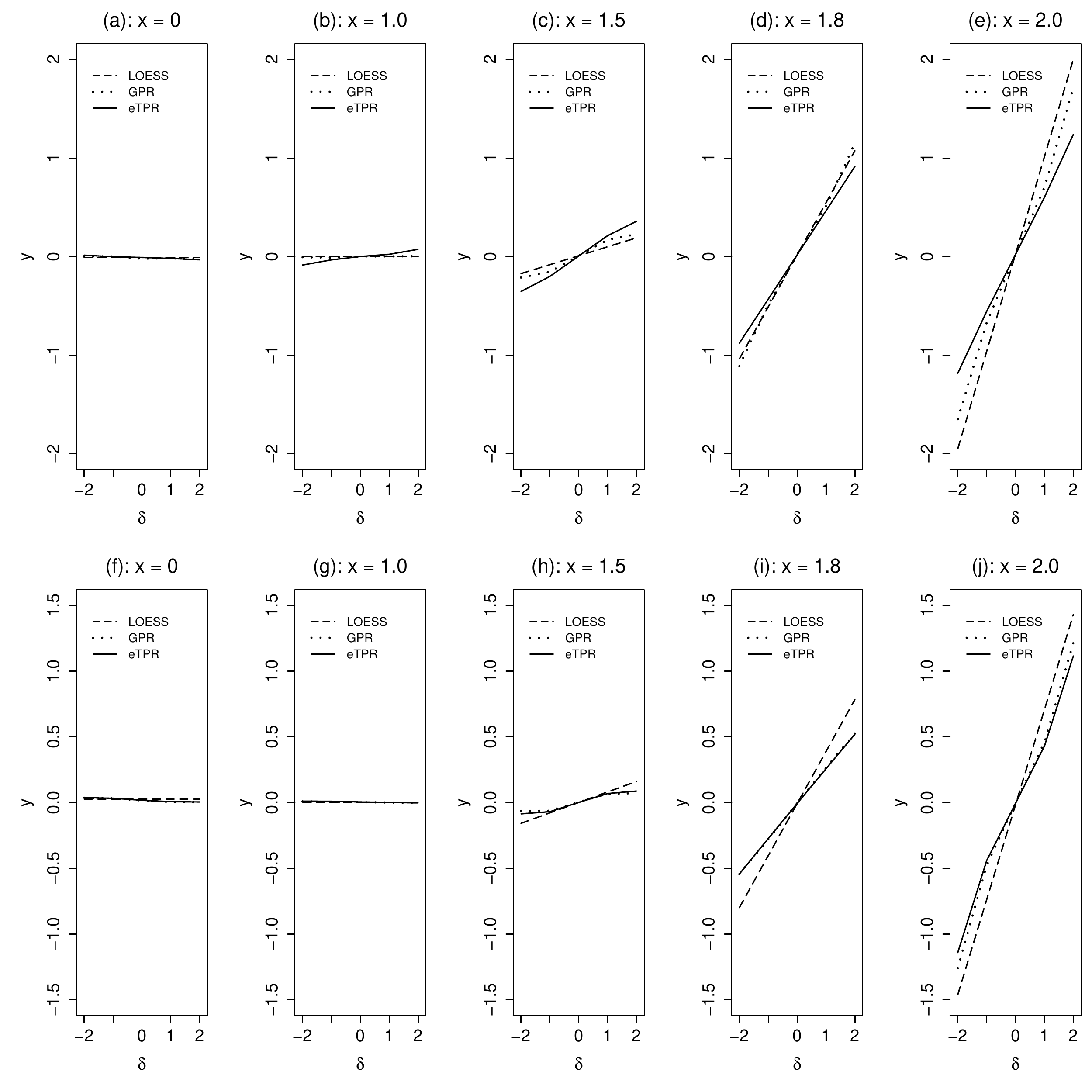}
\end{center}
\caption{Predicted values at the data points $x\in\{0.0,~1.0,~1.5,~1.8,~2.0%
\} $ with constant disturbance at the point 2.0 for sample sizes 10
(sub-figures a-g) and 50 (sub-figures f-j), where dashed, doted and solid
lines respectively represent predictions from the LOESS, GPR and eTPR
methods. }
\label{fig3}
\end{figure}

Instead of random disturbance, a constant disturbance is added to the last
data point 2 in training data. Let $y_{n}^{\ast }=y_{n}+\delta $ where $%
\delta =-2,-1,0,1$ and $2$. Then predicted values $\hat{y}(u)$ at data
points $u=0,1.0,1.5,1.8$ and $2.0$ are calculated by the LOESS, GPR and
eTPR. In each data point, Figure \ref{fig3} shows an average value of
predictions from 500 simulated data, where the true function is 0, and
dashed, doted and solid lines respectively represent average prediction from
the LOESS, GPR and eTPR methods. In a small sample with $n=10$, Figure \ref{fig3}(a-e) shows that the
predictions from the LOESS and the GPR methods tend to be shrunken more at
dense region $0\leq x\leq 1.5$, while those from the eTPR method are
shrunken heavily at sparse region $1.5<x\leq 2.0$. For moderate sample size $%
n=50$, it follows from Figure \ref{fig3}(f-j) that the eTPR behaves like the GPR, and the eTPR shrinks data points 1.8
and 2.0 more heavily than the LOESS.


\noindent \textbf{Robust property against outliers in output space}

We generate the data $y_{i}$ under five process models as follows:\newline
(1) $f\sim GP(h,k)$, $\epsilon \sim N(0,\phi )$ and $%
\mbox{\boldmath
${\beta}$}=(0.1,~0.01,~10,~0.01)=\mbox{{\boldmath
${\beta}$}$_{1}$}$;\newline
(2) $f\sim GP(h,k)$, $\epsilon \sim N(0,\phi )$ and $%
\mbox{\boldmath
${\beta}$}=(0.2,~0.2,~10,~0.1)=\mbox{{\boldmath ${\beta}$}$_{2}$}$;\newline
(3) $f\sim GP(h,k)$, $\epsilon \sim \phi t_{2}$ and $%
\mbox{\boldmath
${\beta}$}=\mbox{{\boldmath ${\beta}$}$_{1}$}$;\newline
(4) $f\sim GP(h,k)$, $\epsilon \sim \phi t_{2}$ and $%
\mbox{\boldmath
${\beta}$}=\mbox{{\boldmath ${\beta}$}$_{2}$}$;\newline
(5) $f\sim ETP(2,2,h,k)$, $\epsilon \sim ETP(2,2,0,\tilde{k})$ and $%
\mbox{\boldmath ${\beta}$}$=(0.1,~0.02,~10,~0.02)$=%
\mbox{{\boldmath
${\beta}$}$_{3}$}$,\newline
where $h(x)=\cos (x)$ or $\cos (2x)$ for $x\in (0,3)$. Let $S$ be a set of
40 points evenly spaced in the interval (0, 3). We randomly take $n=10$ data
points from $S$ as the training data set, and the rest as test data set.
Values of mean squared error, $MSE=\sum_{j=1}^{m}(\hat{f}%
(x_{i}^{\ast })-h(x_{i}^{\ast }))^{2}/m$ for test data points $\{x_{i}^{\ast
},i=1,...,m\}$, are computed by using the LOESS, GPR and eTPR methods. Table \ref{tab2-1} shows the
MSEs for Cases (1) and (2), where the data are generated from
GPR models. We can see that all three methods work similarly. Now we consider cases for model
misspecifications and/or the presence of outliers. For Cases (1), (2) and
(5), one data point is randomly selected from the training data set and is
added with a $t_{1}$ error to study robustness of the proposed methods. Now
Cases (1) and (2) have outliers, Cases (3) and (4) have non-normal errors and
Case (5) has both. We see from Table  \ref{tab2} that the eTPR method gives BLUPs with much
smaller MSE than the LOESS and GPR methods.

We also study robustness of BLUPs from the eTPR model with multivariate
covariates. We consider $h_{1}(\mbox{\boldmath ${x}$}%
)=0.5x_{1}|x_{1}|^{1/3}-3\cos (x_{2})+\log (x_{3})$ and $h_{2}(%
\mbox{\boldmath ${x}$})=0.2x_{1}^{3}+\sin (x_{2})+0.2\exp (x_{3})$ with $%
\mbox{\boldmath ${x}$}=(x_{1},x_{2},x_{3})^{T}$. In this case, parameter $%
\mbox{\boldmath ${\beta}$}$ is $(\phi ,\theta _{0},%
\mbox{{\boldmath
${\theta}$}$_{1}$},\mbox{{\boldmath ${\theta}$}$_{2}$})$ with $%
\mbox{{\boldmath
${\theta}$}$_{1}$}=(\theta _{11},\theta _{12},\theta _{13})$ and $%
\mbox{{\boldmath ${\theta}$}$_{2}$}=(\theta _{21},\theta _{22},\theta _{23})$%
. To generate the data, we follow the previous five process models, but $%
\mbox{\boldmath ${\beta}$}_{1}=(0.1,~0.01,~10,~10,~10,~0.01,~0.01,~0.01)$, $%
\mbox{\boldmath ${\beta}$}_{2}$=(0.2,~0.05,~10, ~10,~10,~0.05, ~0.05,~0.05),
and $\mbox{\boldmath
${\beta}$}_{3}=$(0.1,~0.02,~10,~10, 10,~0.02,~0.02,~0.02). Let $S_{1}$, $%
S_{2}$ and $S_{3}$ be sets of 80 points evenly spaced in the intervals (-2,
2), (0, 3) and (1, 2), respectively. We take $n=30$ random points as
training data and the remaining $m=50$ points as test data. For Cases (1),
(2) and (5), two data points are randomly selected from the training data
set and are added with two independent $t_{1}$ errors. Table \ref{tab3}
presents MSE results. Again, BLUPs from the eTPR method is better than those
from the LOESS and GPR methods. As the number of covariates increases, the
LOESS has the worst MSE.

\begin{table}[th]
\caption{Mean squared errors of prediction results and their standard
deviation (in parentheses) by the LOESS, GPR and eTPR methods with function $%
h(x)=\cos (x)$ and $\cos (2x)$ and Cases (1) and (2) of data generation.}
\label{tab2-1}\tabcolsep=5pt \fontsize{10}{16}\selectfont
 \vskip 0pt
\par
\begin{center}
\begin{tabular}{ccccc}
function & Model & LOESS & GPR & eTPR \\ \hline
cos(x) & (1) & 0.450(0.492) & 0.455(0.486) & 0.450(0.481) \\
& (2) & 0.587(0.620) & 0.556(0.554) & 0.549(0.555) \\ \cline{2-5}
cos(2x) & (1) & 0.459(0.499) & 0.514(0.498) & 0.502(0.486) \\
& (2) & 0.595(0.629) & 0.627(0.560) & 0.643(0.603) \\ \hline
\end{tabular}
\end{center}
\end{table}

\begin{table}[!ht]
\caption{Mean squared errors of prediction results and their standard
deviation (in parentheses) by the LOESS, GPR and eTPR methods with function $%
h(x)=\cos(x)$ and $\cos(2x)$.}
\label{tab2}\tabcolsep=5pt \fontsize{10}{16}\selectfont
 \vskip 0pt
\par
\begin{center}
\begin{tabular}{ccccc}
function & Model & LOESS & GPR & eTPR \\ \hline
cos(x) & (1) & 1.765(7.831) & 0.812(1.634) & 0.569(0.743) \\
& (2) & 1.938(7.778) & 0.901(1.612) & 0.666(1.030) \\
& (3) & 0.887(4.014) & 0.641(1.117) & 0.611(1.027) \\
& (4) & 1.137(1.461) & 0.886(1.880) & 0.808(1.495) \\
& (5) & 2.130(8.511) & 0.838(3.956) & 0.355(0.458) \\ \cline{2-5}
cos(2x) & (1) & 1.771(7.827) & 0.912(1.484) & 0.665(0.777) \\
& (2) & 1.943(7.775) & 0.974(1.447) & 0.720(0.751) \\
& (3) & 0.891(3.998) & 0.741(1.132) & 0.684(1.121) \\
& (4) & 1.139(1.453) & 0.996(1.867) & 0.900(1.653) \\
& (5) & 2.030(8.119) & 0.752(3.086) & 0.447(0.465) \\ \hline
\end{tabular}
\end{center}
\end{table}

\begin{table}[!ht]
\caption{Mean squared errors of prediction results and their standard
deviation (in parentheses) by the LOESS, GPR and eTPR methods with
multivariate mean functions $h(\mbox{\boldmath ${x}$})=h_1(
\mbox{\boldmath
${x}$})$ and $h_2(\mbox{\boldmath ${x}$})$.}
\label{tab3}\tabcolsep=5pt \fontsize{10}{16}\selectfont
 \vskip 0pt
\par
\begin{center}
\begin{tabular}{ccccc}
$h(\mbox{\boldmath ${x}$})$ & Model & LOESS & GPR & eTPR \\ \hline
$h_1(\mbox{\boldmath ${x}$})$ & (1) & 1.086(5.000) & 0.468(1.669) &
0.453(1.761) \\
& (2) & 1.237(5.001) & 0.774(1.771) & 0.768(1.728) \\
& (3) & 0.654(0.425) & 0.272(1.350) & 0.207(0.445) \\
& (4) & 0.884(0.829) & 0.738(2.684) & 0.632(0.941) \\
& (5) & 1.212(4.255) & 0.761(2.247) & 0.640(1.452) \\ \cline{2-5}
$h_2(\mbox{\boldmath ${x}$})$ & (1) & 1.120(4.471) & 0.490(2.088) &
0.306(0.837) \\
& (2) & 1.262(4.489) & 0.813(2.100) & 0.615(0.882) \\
& (3) & 0.753(0.437) & 0.289(0.580) & 0.234(0.297) \\
& (4) & 0.987(0.872) & 0.745(1.075) & 0.634(0.653) \\
& (5) & 1.312(4.405) & 0.718(1.781) & 0.528(0.788) \\ \hline
\end{tabular}
\end{center}
\end{table}

\vskip 10pt

\subsection{Real examples}

\vskip 10pt

\begin{figure}[h!]
\begin{center}
\includegraphics[height=0.6\textwidth,width=0.7\textwidth]{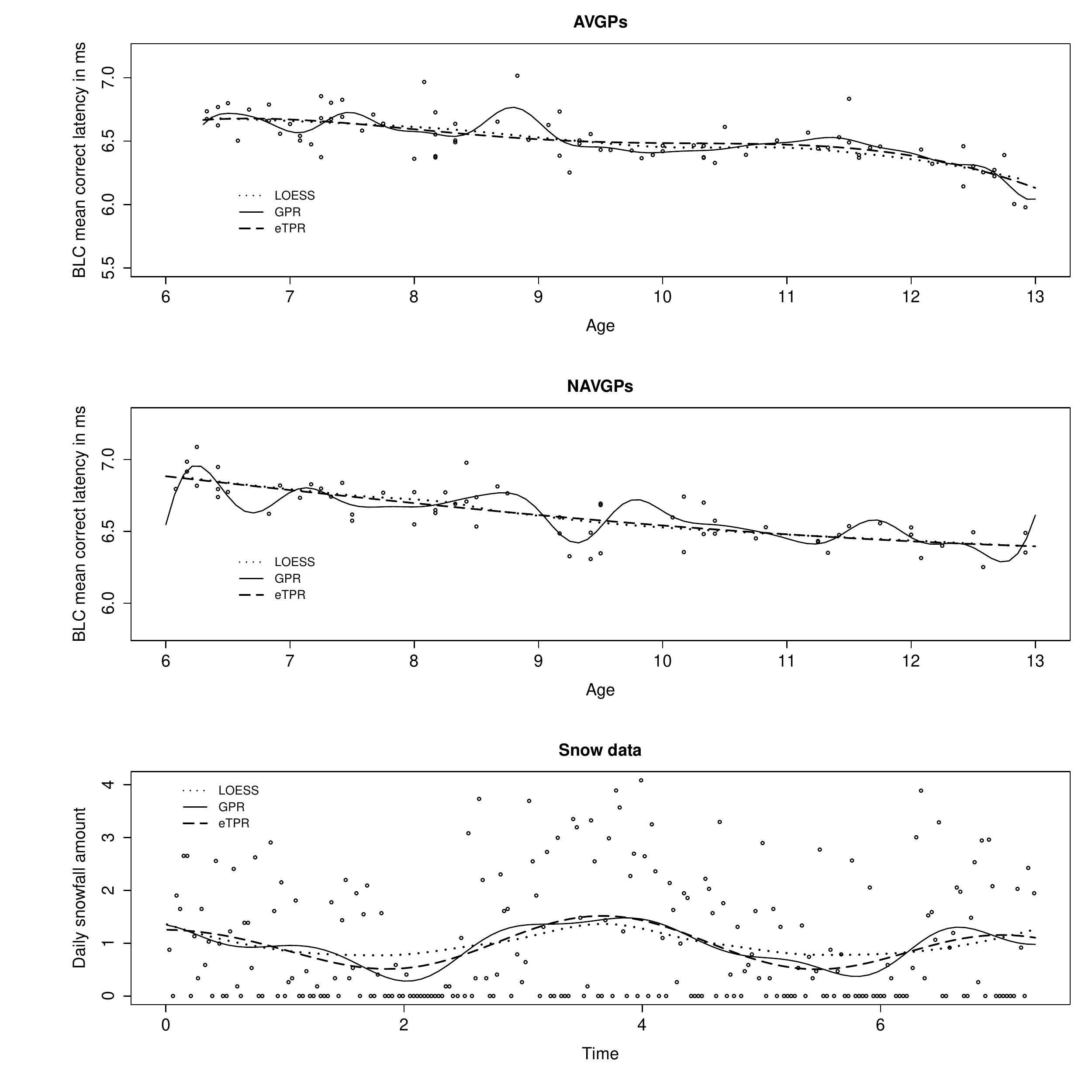}
\end{center}
\caption{Prediction curves from the LOESS, GPR and eTPR methods for 2 groups
of the BLC data and snow data, where circles represent data points, and
solid, dotted and dashed lines stand for predictions from the GPR, LOESS and
eTPR, respectively. }
\label{fig4}
\end{figure}

\begin{figure}[h!]
\begin{center}
\includegraphics[height=0.4\textwidth,width=0.7\textwidth]{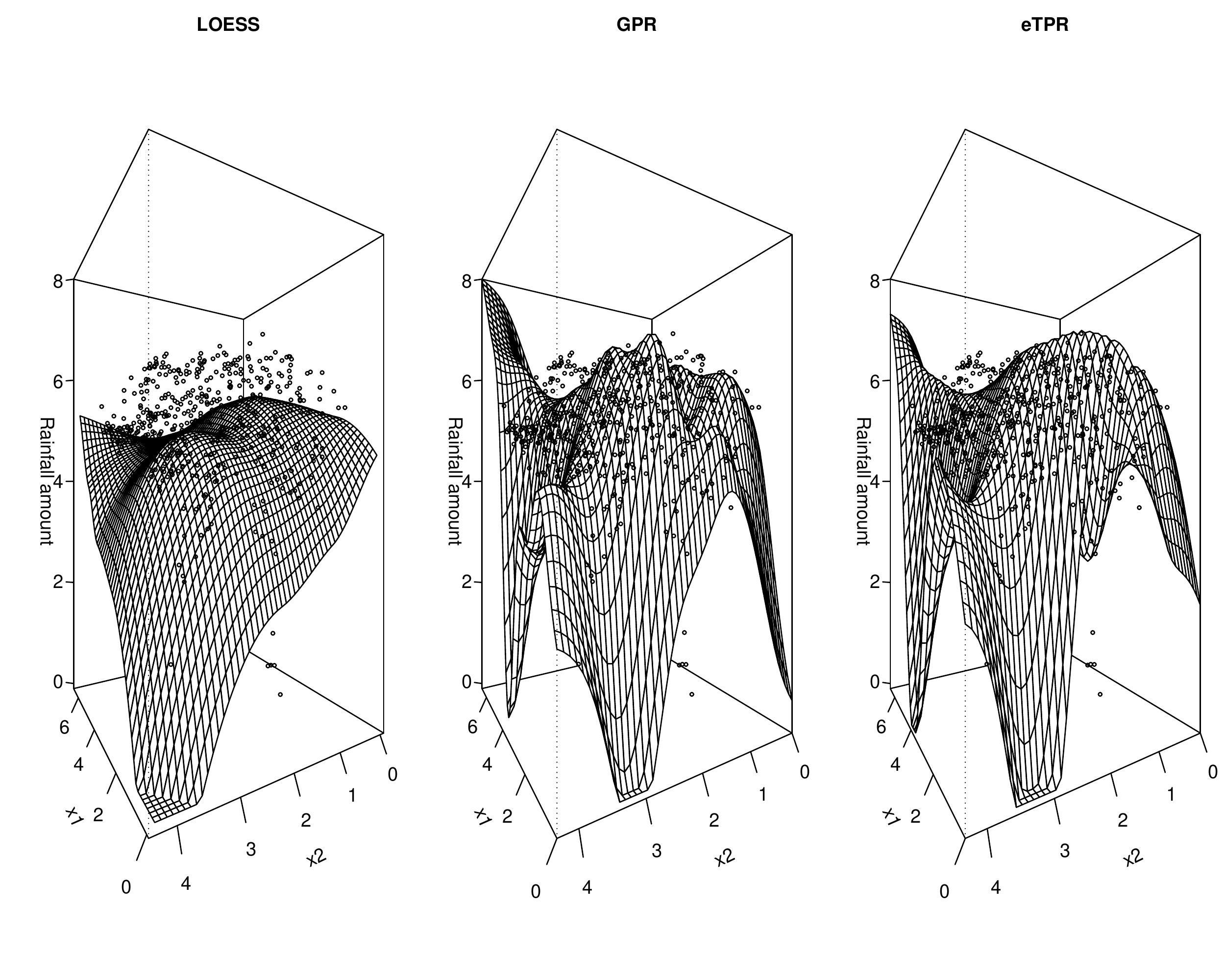}
\end{center}
\caption{Prediction surfaces from the LOESS, GPR and eTPR methods for
spatial data, where circles represent data points. }
\label{fig5}
\end{figure}

The eTPR model (\ref{assum1}) is applied to three data sets. Executive function research
data come from the study in children with Hemiplegic Cerebral Palsy
consisting of 84 girls and 57 boys from primary and secondary schools. These
students were subdivided into two groups: the action video game players
group (AVGPs) (56\%) and the non action video game players group (NAVGPs)
(44\%). In this study, Big/Little Circle (BLC) mean correct latency is
investigated as age of children: for more details of this data set, see Xu
\textit{et al}. (2015). Before applying the proposed methods, we take
logarithm of Big/Little Circle (BLC) mean correct latency. Figure \ref{fig4}
presents prediction curves for 2 groups: AVGPs and NAVGPs, where circles
represent observed data points, and solid line, dashed line and dotted line
stand for predictions from the GPR, eTPR and LOESS methods, respectively. We
can see prediction curves from the LOESS and eTPR methods are more smooth
than those from the GPR method.

Whistler snowfall data contain daily snowfall amounts
in Whistler for the years 2010 and 2011, and can be downloaded at
http://www.climate.weather office.ec.gc.ca. Response for snow data is
logarithm of (daily snowfall amount+1) and covariate is time. From Figure
\ref{fig4}, we can see that predicted curve from the LOESS is the most
smooth, while that from the GPR is the least smooth.

For spatial interpolation data, rainfall measurements at 467 locations were
recorded in Switzerland on 8 May 1986, and can be found at
http://www.ai-geostats.org under SIC97. Spatial interpolation data has
response, logarithms of (rainfall amount+1), and two covariates for
coordinates of location. Prediction surfaces of spatial data are presented
in Figure \ref{fig5}. We can see again that the LOESS surface is the most
smooth while the GPR one is the least smooth.

We randomly select 80\% observation as training data and compute prediction
errors for the remaining data points (i.e. the test data). This procedure is
repeated 500 times. Table \ref{tab4} presents mean prediction errors of
these 3 data sets. We can see that the LOESS is the best in BLC-AVGPs, while
it is the worst in the snow and spatial data particularly for the latter
which includes multivariate predictors. Overall, the eTPR is the best in
prediction.

\begin{table}[!ht]
\caption{Prediction errors and their standard deviation (in parentheses) for
the 3 real data sets by the LOESS, GPR and eTPR methods.}
\label{tab4}\tabcolsep=5pt \fontsize{10}{16}\selectfont
 \vskip 0pt
\par
\begin{center}
\begin{tabular}{cccc}
Data & LOESS & GPR & eTPR \\ \hline
BLC-AVGPs & 0.022(0.010) & 0.031(0.015) & 0.024(0.010) \\
BLC-NAVGPs & 0.017(0.006) & 0.025(0.033) & 0.016(0.006) \\
Snow & 1.133(0.097) & 1.117(0.102) & 1.116(0.101) \\
Spatial & 0.524(0.121) & 0.210(0.086) & 0.204(0.089) \\ \hline
\end{tabular}
\end{center}
\end{table}

\section{Concluding remarks}

Advantages of a GPR model include that it offers a nonparametric
regression model for data with multi-dimensional covariates, the
specification of covariance kernel enables to accommodate a wide class of
nonlinear regression functions, and it can be applied to analyze many
different types of data including functional data. In this paper, we extended the GPR
model to the eTPR model. The latter inherits almost all the good
features for the GPR, and additionally it provides robust BLUP procedures in the
presence of outliers in both input and output spaces. Numerical studies show
that the eTPR is overall the best in prediction among the methods considered.

\section*{Appendix}

Let $\mbox{\boldmath${\Sigma}$}$ be
an $n\times n$ symmetric and positive definite matrix, $\mbox{\boldmath${%
\mu}$}\in {R}{^{n}} $, $\nu >0$ and $\omega >0$. In this paper, $\mbox{\boldmath${Z}$}\sim
EMTD(\nu ,\omega ,\mbox{\boldmath${\mu}$},\mbox{\boldmath${\Sigma}$})$ means that a
random vector {$\mbox{\boldmath${Z}$}\in {R}{^{n}}$} has the
density function,
\begin{equation*}
p(z)=|2\pi \omega \mbox{\boldmath${\Sigma}$}|^{-1/2}\frac{\Gamma (n/2+\nu )}{%
\Gamma (\nu )}\left( 1+\frac{\mbox{{\boldmath${(z-\mu)}$}$^{T}$}%
\mbox{{\boldmath${\Sigma}$}$^{{-1}}$}\mbox{\boldmath${(z-\mu)}$}}{2\omega }%
\right) ^{-(n/2+\nu )},
\end{equation*}
where $\Gamma (\cdot )$ is the gamma function.


We may construct an EMTD via a double hierarchical generalized linear model
(Lee and Nelder, 2006) as follows: \newline
\noindent \textbf{Lemma 1} \textit{If
\begin{equation*}
\mbox{\boldmath${Z}$}|r\sim N(\mbox{\boldmath${\mu}$},r%
\mbox{\boldmath
${\Sigma}$}),~~r\sim \mathrm{IG}(\nu ,\omega ),
\end{equation*}
where $\mathrm{IG}(\nu ,\omega )$ stands for an inverse gamma distribution
with the density function
\begin{equation*}
g(r)=\frac{1}{\Gamma (\nu )}(\frac{\omega }{r})^{\nu +1}\frac{1}{\omega }%
\exp {(-\frac{\omega }{r}),}
\end{equation*}
then, marginally $\mbox{\boldmath${Z }$}\sim EMTD(\nu ,\omega ,%
\mbox
{\boldmath${\mu}$},\mbox{\boldmath${\Sigma}$})$.}

\vskip 10pt \noindent \textbf{Proof}: From the construction
of $\mbox{\boldmath ${Z}$}$, we have
\begin{align}
p(\mbox{\boldmath ${z}$})&=\int_0^\infty p(\mbox{\boldmath ${z}$}|r)g(r)dr
\notag \\
&=\int_0^\infty |2\pi r\mbox{\boldmath ${\Sigma}$}|^{-1/2} \exp{(-\frac{(%
\mbox{\boldmath ${z}$}-\mbox{\boldmath ${\mu}$})^T%
\mbox{{\boldmath
${\Sigma}$}$^{-1}$}(\mbox{\boldmath ${z}$}-\mbox{\boldmath ${\mu}$})}{2r})}
\frac{1}{\Gamma(\nu)}(\frac{\omega}{r})^{\nu+1}\frac{1}{\omega}\exp{%
(-\omega/r)} dr  \notag \\
&=\int_0^\infty |2\pi \omega \mbox{\boldmath ${\Sigma}$}|^{-1/2} \frac{1}{%
\Gamma(\nu)} (\frac{\omega}{r})^{n/2+\nu-1} \exp{(-\frac{2\omega+(%
\mbox{\boldmath ${z}$}-\mbox{\boldmath ${\mu}$})^T%
\mbox{{\boldmath
${\Sigma}$}$^{-1}$}(\mbox{\boldmath ${z}$}-\mbox{\boldmath ${\mu}$})}{2r})} d%
\frac{\omega}{r}  \notag \\
&=|2\pi \omega \mbox{\boldmath ${\Sigma}$}|^{-1/2} \frac{\Gamma(n/2+\nu)}{%
\Gamma(\nu)}\left(1+\frac{(\mbox{\boldmath ${z}$}-\mbox{\boldmath ${\mu}$})^T%
\mbox{{\boldmath ${\Sigma}$}$^{-1}$}(\mbox{\boldmath ${z}$}-%
\mbox{\boldmath
${\mu}$})}{2\omega}\right)^{-(n/2+\nu)},  \notag
\end{align}
which is the density function of EMTD.$\sharp$

Properties of EMTD are as follows.

\noindent\textbf{Lemma 2} \textit{Let $\mbox{\boldmath${Z }$}\sim
EMTD(\nu,\omega,\mbox{\boldmath${\mu}$},\mbox{\boldmath${\Sigma}$}).$}

\begin{itemize}
\item[(i)]  \textit{If $\omega /\nu \rightarrow \lambda >0$ as $\nu
\rightarrow \infty $, then $\lim_{\nu \rightarrow \infty }EMTD(\nu ,\omega ,%
\mbox{\boldmath${\mu}$},\mbox{\boldmath${\Sigma}$})=N(\mu ,\lambda %
\mbox{\boldmath${\Sigma}$}).$ }

\item[(ii)]  \textit{{For any matrix $\mbox{\boldmath${A}$}\in R^{l\times n}$
with rank $l\leq n$, }} $\mathit{{\mbox{\boldmath${A }$}\mbox{\boldmath${Z}$}%
\sim EMTD(\nu ,\omega ,\mbox{\boldmath${A }$}\mu ,\mbox{\boldmath${A }$}%
\mbox{\boldmath${\Sigma}$}\mbox{{\boldmath${A}$}$^{T}$})}.\ }$

\item[(iii)]  \textit{Let $\mbox{\boldmath${Z}$}$ be partitioned as ${(%
\mbox{{\boldmath${Z}$}$_{1}^{T}$},\mbox{{\boldmath${Z}$}$_{2}^{T}$})}^{T}$
with lengths $n_{1}$ and $n_{2}=n-n_{1}$, and $\mbox{\boldmath${\mu}$}$ and $%
\mbox{\boldmath${\Sigma}$}$ have the corresponding partitions as $%
\mbox{\boldmath${\mu}$}=(\mbox{{\boldmath${\mu}$}$_{1}^{T}$},%
\mbox
{{\boldmath${\mu}$}$_{2}^{T}$})^{T}$ and $\mbox{\boldmath${\Sigma}$}=\left(
\begin{array}{cc}
\mbox{{\boldmath${\Sigma}$}$_{{11}}$} &
\mbox{{\boldmath
${\Sigma}$}$_{{12}}$} \\
\mbox{{\boldmath${\Sigma}$}$_{{12}}^{T}$} &
\mbox{{\boldmath
${\Sigma}$}$_{{22}}$}
\end{array}
\right) $. Then,}
\begin{align}
& \mathit{\mbox{{\boldmath${Z}$}$_{1}$}\sim EMTD(\nu ,\omega ,%
\mbox{{\boldmath${\mu}$}$_{1}$},\mbox{{\boldmath${\Sigma}$}$_{{11}}$}),}
\notag \\
& \mathit{\mbox{{\boldmath${Z}$}$_{2}$}|\mbox{{\boldmath${Z}$}$_{1}$}=%
\mbox{{\boldmath${z}$}$_{1}$}\sim EMTD(\nu ^{\ast },\omega ^{\ast },%
\mbox{\boldmath${\mu}$}^{\ast },\mbox{\boldmath${\Sigma}$}^{\ast })},  \notag
\end{align}
\textit{with $\nu ^{\ast }=n_{1}/2+\nu $, $\omega ^{\ast }=n_{1}/2+\omega $,
} $\mbox {{\boldmath${\mu}$}$^{*}$}=\mbox{{\boldmath${\Sigma}$}$_{{12}}^{T}$}%
\mbox{{\boldmath${\Sigma}$}$_{{11}}^{{-1}}$}(\mbox{{\boldmath${z}$}$_{1}$}-%
\mbox{{\boldmath${\mu}$}$_{1}$})+\mbox{{\boldmath${\mu}$}$_{2}$},$ $%
\mbox{{\boldmath${\Sigma}$}$^{*}$}=(2\omega +(\mbox{{\boldmath${z}$}$_{1}$}-%
\mbox{{\boldmath${\mu}$}$_{1}$})^{T}%
\mbox{{\boldmath
${\Sigma}$}$_{{11}}^{{-1}}$}(\mbox{{\boldmath${z}$}$_{1}$}-%
\mbox{{\boldmath
${\mu}$}$_{1}$}))\mbox{{\boldmath${\Sigma}$}$_{{22\cdot1}}$}/({2\omega +n_{1}%
}),$ \textit{and $\mbox{{\boldmath${\Sigma}$}$_{{22\cdot1}}$}=%
\mbox{{\boldmath
${\Sigma}$}$_{{22}}$}-\mbox{{\boldmath
${\Sigma}$}$_{{12}}^{T}$}\mbox{{\boldmath${\Sigma}$}$_{{11}}^{{-1}}$}%
\mbox{{\boldmath${\Sigma}$}$_{{12}}$}$. This gives } $E(\mbox{{%
\boldmath${Z}$}$_{2}$}|\mbox{{\boldmath${Z}$}$_{1}$})=%
\mbox
{{\boldmath${\mu}$}$^{*}$}\text{\textit{\ and }}Cov(\mbox{{%
\boldmath${Z}$}$_{2}$}|\mbox{{\boldmath${Z}$}$_{1}$})={\omega ^{\ast }}%
\mbox{{\boldmath${\Sigma}$}$^{*}$}/({\nu ^{\ast }-1}).$

\item[(iv)]  \textit{Let $r$ be a random effect in Proposition 1. Then, $r|%
\mbox{\boldmath${Z}$}\sim \mathrm{IG}(\tilde{\nu},\tilde{\omega})$ with $%
\tilde{\nu}=n/2+\nu ,$ $\tilde{\omega}=\omega +(\mbox{\boldmath${Z}$}-%
\mbox{\boldmath${\mu}$})^{T}\mbox{{\boldmath
${\Sigma}$}$^{{-1}}$}(\mbox{\boldmath${Z}$}-\mbox{\boldmath${\mu}$})/2$ and
\begin{align}
E(r|\mbox{\boldmath${Z}$})=\frac{2\omega +(\mbox{\boldmath${Z}$}-%
\mbox{\boldmath${\mu}$})^{T}\mbox{{\boldmath${\Sigma}$}$^{{-1}}$}(%
\mbox{\boldmath${Z}$}-\mbox{\boldmath${\mu}$})}{n+2\nu -2},  \notag \\
Var(r|\mbox{\boldmath${Z}$})=\frac{(2\omega +(\mbox{\boldmath${Z}$}-%
\mbox{\boldmath${\mu}$})^{T}\mbox{{\boldmath${\Sigma}$}$^{{-1}}$}(%
\mbox{\boldmath${Z}$}-\mbox{\boldmath${\mu}$}))^{2}}{(n+2\nu -2)^{2}(n/2+\nu
-2)}.  \notag
\end{align}
}
\end{itemize}

\vskip10pt \noindent \textbf{Proof}: The conclusions (i),
(ii) and $\mbox{{\boldmath ${Z}$}$_{1}$}\sim EMTD(\nu ,\omega ,%
\mbox{\boldmath ${\mu}$}_{1},\mbox{{\boldmath ${\Sigma}$}$_{11}$})$ in (iii)
are easily obtained by the definition of EMTD and Lemma 1. Now we only
prove that  $\mbox{{\boldmath ${Z}$}$_{2}$}|\mbox{{\boldmath ${Z}$}$_{1}$}\sim
EMTD(\nu ^{\ast },\omega ^{\ast },\mbox{{\boldmath ${\mu}$}$^{*}$},%
\mbox{{\boldmath ${\Sigma}$}$^{*}$})$. Let $a _{1}=(%
\mbox{{\boldmath
${z}$}$_{1}$}-\mbox{{\boldmath ${\mu}$}$_{1}$})^{T}%
\mbox{{\boldmath
${\Sigma}$}$_{11}^{-1}$}(\mbox{{\boldmath ${z}$}$_{1}$}-%
\mbox{{\boldmath
${\mu}$}$_{1}$})$ and $a _{2}=(\mbox{{\boldmath ${z}$}$_{2}$}-%
\mbox{{\boldmath ${\mu}$}$^{*}$})^{T}\mbox{{\boldmath ${\Sigma}$}$^{*-1}$}(%
\mbox{{\boldmath ${z}$}$_{2}$}-\mbox{{\boldmath ${\mu}$}$^{*}$})$, then $a
_{1}+a _{2}=(\mbox{\boldmath ${z}$}-\mbox{\boldmath ${\mu}$})^{T}%
\mbox{{\boldmath ${\Sigma}$}$^{-1}$}(\mbox{\boldmath ${z}$}-%
\mbox{\boldmath
${\mu}$})$. We have
\begin{align}
&p(\mbox{{\boldmath ${z}$}$_{2}$}|\mbox{{\boldmath ${z}$}$_{1}$}) =\frac{p(%
\mbox{\boldmath ${z}$})}{p(\mbox{{\boldmath ${z}$}$_{1}$})}  \notag \\
=&\frac{|2\pi \omega \mbox{\boldmath ${\Sigma}$}|^{-1/2}\frac{\Gamma
(n/2+\nu )}{\Gamma (\nu )}\left( 1+\frac{a_{1}+a_{2}}{2\omega }\right)
^{-(n/2+\nu )}}{|2\pi \omega \mbox{{\boldmath ${\Sigma}$}$_{11}$}|^{-1/2}%
\frac{\Gamma (n_{1}/2+\nu )}{\Gamma (\nu )}\left( 1+\frac{a_{1}}{2\omega }%
\right) ^{-(n_{1}/2+\nu )}}\propto \left( 1+\frac{a_{2}}{2\omega +a_{1}}%
\right) ^{-(n/2+\nu )},  \notag
\end{align}
which indicates $\mbox{{\boldmath ${Z}$}$_{2}$}|%
\mbox{{\boldmath
${Z}$}$_{1}$}\sim EMTD(\nu ^{\ast },\omega ^{\ast },%
\mbox{{\boldmath
${\mu}$}$^{*}$},\mbox{{\boldmath ${\Sigma}$}$^{*}$})$.

By combining definitions of IG and EMTD, we have
\begin{align}
& p(r|\mbox{\boldmath ${Z}$})=\frac{p(\mbox{\boldmath ${Z}$}|r)g(r)}{p(%
\mbox{\boldmath ${Z}$})}  \notag \\
& =\frac{1}{\Gamma (n/2+\nu )}\frac{1}{\omega +(\mbox{\boldmath ${Z}$}-%
\mbox{\boldmath ${\mu}$})^{T}\mbox{{\boldmath ${\Sigma}$}$^{-1}$}(%
\mbox{\boldmath ${Z}$}-\mbox{\boldmath ${\mu}$})/2}\left( \frac{\omega +(%
\mbox{\boldmath ${Z}$}-\mbox{\boldmath ${\mu}$})^{T}%
\mbox{{\boldmath
${\Sigma}$}$^{-1}$}(\mbox{\boldmath ${Z}$}-\mbox{\boldmath ${\mu}$})/2}{r}%
\right) ^{n/2+\nu +1}  \notag \\
& \exp {\left( -\frac{\omega +(\mbox{\boldmath ${Z}$}-%
\mbox{\boldmath
${\mu}$})^{T}\mbox{{\boldmath ${\Sigma}$}$^{-1}$}(\mbox{\boldmath ${Z}$}-%
\mbox{\boldmath ${\mu}$})/2}{r}\right) },  \notag
\end{align}
which indicates (iv) holds in this Lemma.$\sharp$

\vskip10pt \noindent \textbf{Proof of Proposition 1}: Proposition 1 can be
easily proved by using Lemma 2, so omitted here.$\sharp$

\vskip 10pt \noindent \textbf{Marginal likelihood derivatives:}

We know that $\mbox{{\boldmath ${y}$}$_{n}$}|\mbox{{\boldmath ${X}$}$_{n}$}\sim
EMTD(\nu ,\nu-1 ,0,\mbox{{\boldmath
${\tilde{\Sigma}}$}$_{n}$})$. For given $\nu $, the marginal log-likelihood
of $\mbox{\boldmath ${\beta}$}$ is
\begin{align}
l(\mbox{\boldmath ${\beta}$};\nu)=& -\frac{n}{2}\log (2\pi (\nu-1) )-\frac{1%
}{2}\log |\mbox{{\boldmath ${\tilde{\Sigma}}$}$_{n}$}|{-(\frac{n}{2}+\nu )}%
\log \left( 1+\frac{S}{2(\nu-1) }\right)  \notag \\
& +\log (\Gamma (\frac{n}{2}+\nu ))-\log (\Gamma (\nu )),  \notag
\end{align}
where $S=\mbox{{\boldmath ${y}$}$_{n}^{T}$}\mbox{{\boldmath
${\tilde{\Sigma}}$}$_{n}^{-1}$}\mbox{{\boldmath ${y}$}$_{n}$}$. The
derivative with respect to $\mbox{\boldmath ${\beta}$}$ is
\begin{equation}
\frac{\partial l(\mbox{\boldmath ${\beta}$};\nu ,(\nu-1) )}{\partial %
\mbox{\boldmath ${\beta}$}}=\frac{1}{2}Tr\left( \Big(\frac{n+2\nu }{2(\nu-1)
+S}\mbox{\boldmath ${\alpha}$}\mbox{{\boldmath ${\alpha}$}$^{T}$}-%
\mbox{{\boldmath ${\tilde{\Sigma}}$}$_{n}^{-1}$}\Big)\frac{\partial %
\mbox{{\boldmath ${\tilde{\Sigma}}$}$_{n}$}}{\partial
\mbox{\boldmath
${\beta}$}}\right) ,  \label{score-beta}
\end{equation}
where $\mbox{\boldmath ${\alpha}$}=%
\mbox{{\boldmath
${\tilde{\Sigma}}$}$_{n}^{-1}$}\mbox{{\boldmath ${y}$}$_{n}$}$.

Estimates of parameters $\mbox{\boldmath ${\beta}$}$ can be learned by using
gradient based methods. And variances of the estimates can be estimated by
computing the second derivatives of $l(\mbox{\boldmath
${\beta}$};\nu)$ on $\mbox{\boldmath ${\beta}$}$ as follows,
\begin{align}
&\frac{\partial^2 l(\mbox{\boldmath ${\beta}$};\nu)}{\partial %
\mbox{\boldmath ${\beta }$} \partial \mbox{\boldmath ${\beta}$}}=\frac{1}{2}%
Tr\left(\Big(\frac{n+2\nu}{2(\nu-1)+S}\mbox{\boldmath ${\alpha}$}%
\mbox{{\boldmath ${\alpha}$}$^{T}$}-%
\mbox{{\boldmath
${\tilde{\Sigma}}$}$_{n}^{-1}$}\Big)\Big(\frac{\partial ^2
\mbox{{\boldmath
${\tilde{\Sigma}}$}$_{n}$}}{\partial\mbox{\boldmath ${\beta }$} \partial%
\mbox{\boldmath ${\beta}$}}-\frac{\partial
\mbox{{\boldmath
${\tilde{\Sigma}}$}$_{n}$}}{\partial\mbox{\boldmath ${\beta}$}} %
\mbox{{\boldmath ${\tilde{\Sigma}}$}$_{n}^{-1}$} \frac{\partial %
\mbox{{\boldmath ${\tilde{\Sigma}}$}$_{n}$}}{\partial%
\mbox{\boldmath
${\beta}$}}\Big)\right)  \notag \\
&-\frac{1}{2}Tr\left(\frac{n+2\nu}{2(\nu-1)+S}\mbox{\boldmath ${\alpha}$}%
\mbox{{\boldmath ${\alpha}$}$^{T}$}\frac{\partial
\mbox{{\boldmath
${\tilde{\Sigma}}$}$_{n}$}}{\partial\mbox{\boldmath ${\beta}$}} %
\mbox{{\boldmath ${\tilde{\Sigma}}$}$_{n}^{-1}$} \frac{\partial %
\mbox{{\boldmath ${\tilde{\Sigma}}$}$_{n}$}}{\partial%
\mbox{\boldmath
${\beta}$}}\right)+\frac{1}{2}\frac{n+2\nu}{(2(\nu-1)+S)^2}\left\{Tr\left(%
\mbox{\boldmath ${\alpha}$}\mbox{{\boldmath ${\alpha}$}$^{T}$}\frac{\partial %
\mbox{{\boldmath ${\tilde{\Sigma}}$}$_{n}$}}{\partial%
\mbox{\boldmath
${\beta}$}}\right)\right\}^2.  \notag
\end{align}

\vskip10pt \noindent \textbf{Variance of prediction value $\hat y(%
\mbox{\boldmath ${u}$})$}:

From the hierarchical sampling method described in Lemma 1, we have
\begin{equation*}
\left(
\begin{array}{c}
f \\
\epsilon
\end{array}
\right)\Bigg|r \sim GP\left( \nu,(\nu-1),\left(
\begin{array}{c}
h \\
0
\end{array}
\right) ,\left(
\begin{array}{cc}
rk & 0 \\
0 & r\tilde{k}
\end{array}
\right) \right),~~r\sim \mathrm{IG}(\nu ,(\nu-1) ),
\end{equation*}
which suggests that conditional on $r$, $\mbox{{\boldmath ${y}$}$_{n}$}|f,%
\mbox{{\boldmath ${X}$}$_{n}$}\sim N(f(\mbox{{\boldmath ${X}$}$_{n}$}),r\phi %
\mbox{{\boldmath ${I}$}$_{n}$})$ and marginal distribution of $%
\mbox{{\boldmath ${y}$}$_{n}$}|\mbox{{\boldmath ${X}$}$_{n}$}\sim N(%
\mbox{{\boldmath ${h}$}$_{n}$},r\mbox{{\boldmath ${\tilde\Sigma}$}$_{n}$})$.
For given $r$, it follows from the GPR model that $E(\hat f(%
\mbox{\boldmath
${u}$})|r,\mbox{{\boldmath ${\mathcal{D}}$}$_{n}$})=%
\mbox{{\boldmath
${k}$}$_{{\ve u}}^{T}$}\mbox{{\boldmath
${\tilde{\Sigma}}$}$_{n}^{{-1}}$}(\mbox{{\boldmath${ y}$}$_{n}$}-%
\mbox{{\boldmath${h}$}$_{n}$})+h(\mbox{\boldmath ${u}$})$ and $Var(\hat f(%
\mbox{\boldmath ${u}$})|r,\mbox{{\boldmath
${\mathcal{D}}$}$_{n}$})=r(k(\mbox{\boldmath ${u}$},\mbox{\boldmath ${u}$})-%
\mbox{{\boldmath ${k}$}$_{\ve u}^{T}$}\mbox{{\boldmath${\tilde{%
\Sigma}}$}$_{n}^{-1}$}\mbox{{\boldmath ${k}$}$_{\ve u}$}+\phi)$.
Consequently, we have
\begin{align}
&Var(\hat f(\mbox{\boldmath ${u}$})|\mbox{{\boldmath
${\mathcal{D}}$}$_{n}$})=E((\hat f(\mbox{\boldmath ${u}$}))^2|%
\mbox{{\boldmath ${\mathcal{D}}$}$_{n}$})- (E(\hat f(%
\mbox{{\boldmath
${x}$}$^{*}$})|\mbox{{\boldmath ${\mathcal{D}}$}$_{n}$}))^2  \notag \\
=& E_{r}((Var(\hat f(\mbox{\boldmath ${u}$})|r,%
\mbox{{\boldmath
${\mathcal{D}}$}$_{n}$})+(E(\hat f(\mbox{\boldmath ${u}$})|r,%
\mbox{{\boldmath ${\mathcal{D}}$}$_{n}$}))^2)|%
\mbox{{\boldmath
${\mathcal{D}}$}$_{n}$})- (E(\hat f(\mbox{\boldmath ${u}$})|%
\mbox{{\boldmath
${\mathcal{D}}$}$_{n}$}))^2  \notag \\
=&E_{r}(Var(\hat f(\mbox{\boldmath ${u}$})|r,%
\mbox{{\boldmath
${\mathcal{D}}$}$_{n}$})|\mbox{{\boldmath ${\mathcal{D}}$}$_{n}$})+(E(\hat f(%
\mbox{\boldmath ${u}$})|\mbox{{\boldmath ${\mathcal{D}}$}$_{n}$}))^2-
(E(\hat f(\mbox{\boldmath ${u}$})|\mbox{{\boldmath
${\mathcal{D}}$}$_{n}$}))^2  \notag \\
=&s_0\Big(k(\mbox{\boldmath ${u}$},\mbox{\boldmath ${u}$})-%
\mbox{{\boldmath
${k}$}$_{\ve u}^{T}$}\mbox{{\boldmath${\tilde{\Sigma}}$}$_{n}^{-1}$}%
\mbox{{\boldmath ${k}$}$_{\ve
u}$}+\phi\Big),  \notag
\end{align}
where $s_0={(2\nu-2+(\mbox{{\boldmath
${ y}$}$_{n}$}-\mbox{{\boldmath${h}$}$_{n}$})^{T}%
\mbox{{\boldmath
${\tilde{\Sigma}}$}$_{n}^{-1}$}(\mbox{{\boldmath${ y}$}$_{n}$}-%
\mbox
{{\boldmath${h}$}$_{n}$}))}/{(2\nu +n-2)}$.

\vskip10pt \noindent \textbf{Proof of Proposition 2}:
From (\ref{score-beta}), the score functions of $\mbox{\boldmath
${\beta}$}$ based on the eTPR model is 
\begin{align}
&s_{T}(\mbox{\boldmath ${\beta}$};\mbox{{\boldmath ${y}$}$_{n}$})=\frac{1}{2}%
Tr\left(\Big(\frac{n+2\nu}{2(\nu-1)+\mbox{{\boldmath ${ y}$}$_{n}^{T}$}%
\mbox{{\boldmath ${\tilde{\Sigma}}$}$_{n}^{-1}$}%
\mbox{{\boldmath
${y}$}$_{n}$}}\mbox{{\boldmath ${\tilde{\Sigma}}$}$_{n}^{-1}$}%
\mbox{{\boldmath ${y}$}$_{n}$}\mbox{{\boldmath ${ y}$}$_{n}^{T}$}%
\mbox{{\boldmath ${\tilde{\Sigma}}$}$_{n}^{-1}$}-%
\mbox{{\boldmath
${\tilde{\Sigma}}$}$_{n}^{-1}$}\Big)\frac{\partial
\mbox{{\boldmath
${\tilde{\Sigma}}$}$_{n}$}}{\partial\mbox{\boldmath ${\beta}$}}\right).
\notag
\end{align}
The term $({n+2\nu})/({2(\nu-1)+\mbox{{\boldmath ${y}$}$_{n}^{T}$}%
\mbox{{\boldmath ${\tilde{\Sigma}}$}$_{n}^{-1}$}%
\mbox{{\boldmath
${y}$}$_{n}$}})$ in $s_{T}(\mbox{\boldmath ${\beta}$};\mbox{{\boldmath
${y}$}$_{n}$})$ plays an important role in estimating parameter $%
\mbox{\boldmath ${\beta}$}$. For example, when some observations of
responses have very large value or tend to infinity (outliers), the score $%
s_{T}(\mbox{\boldmath ${\beta}$};\mbox{{\boldmath ${y}$}$_{n}$})$ based on
the eTPR model does not tend to infinity.

Let $\mbox{\boldmath ${T}$}(F_n)=\mbox{{\boldmath ${T}$}$_{n}$}(y_1,...,y_n)$
be an estimate of $\mbox{\boldmath ${\beta}$}$, where $F_n$ is the empirical
distribution of $\{y_1,...,y_n\}$ and $T$ is a functional on some subset of
all distributions. Influence function of $T$ at $F$ (Hampel \textit{et al}., 1986) is
defined as
\begin{align}
IF(y; T, F)=\lim_{t\rightarrow 0}\frac{T((1-t)F+t \delta_y)-T(F)}{t},  \notag
\end{align}
where $\delta_y$ put mass 1 on point $y$ and 0 on others.

For given parameter $\nu$, following Hampel \textit{et al}. (1986) estimator $\hat{%
\mbox{\boldmath ${\beta}$}}$ of $\mbox{\boldmath ${\beta}$}$ has the
influence function
\begin{align}
IF(y; \hat{\mbox{\boldmath ${\beta}$}}, F)=-\left(E\left(\frac{\partial^2 l(%
\mbox{\boldmath ${\beta}$};\nu,(\nu-1))}{\partial
\mbox{\boldmath ${\beta
}$} \partial \mbox{{\boldmath ${\beta}$}$^{T}$}}\right)\right)^{-1}s_{T}(%
\mbox{\boldmath ${\beta}$};y).  \notag
\end{align}
Note that the matrix ${\partial^2 l(\mbox{\boldmath
${\beta}$};\nu)}/{\partial \mbox{\boldmath ${\beta }$} \partial %
\mbox{{\boldmath ${\beta}$}$^{T}$}}$ is bounded according to $%
\mbox{{\boldmath ${y}$}$_{n}$}$, which indicates that the influence function
of $\hat{\mbox{\boldmath ${\beta}$}}$ is bounded under the eTPR model.
Similarly, we can obtain that the score function under the GPR model is unbound,
which leads to unbound influence function of parameter estimate.$\sharp$

\vskip 10pt \noindent \textbf{Proof of the equation (\ref{bspred})}:

From sequential bayesian prediction strategy and Bayes' Theorem, we have
\begin{align}
&\prod_{i=1}^np_{\phi_0,\theta_0}(y_i |\mbox{{\boldmath ${X}$}$_{i}$}, %
\mbox{{\boldmath ${y}$}$_{i-1}$}) =p_{\phi_0,\theta_0}(y_1 |\mbox{{\boldmath
${X}$}$_{1}$}) \prod_{i=2}^n\int_{{\mathcal{F}}} p_{\phi_0}(y_i |f,%
\mbox{{\boldmath
${X}$}$_{i}$}, \mbox{{\boldmath ${y}$}$_{i-1}$})dp(f|%
\mbox{{\boldmath
${X}$}$_{i}$}, \mbox{{\boldmath
${y}$}$_{i-1}$})  \notag \\
=&p_{\phi_0,\theta_0}(y_1 |\mbox{{\boldmath ${X}$}$_{1}$})
\prod_{i=2}^n\int_{{\mathcal{F}}} \frac{p_{\phi_0}(%
\mbox{{\boldmath
${y}$}$_{i}$}|f,\mbox{{\boldmath ${X}$}$_{i}$})dp_{\theta_0}(f)}{\int_{{%
\mathcal{F}}} p_{\phi_0}(\mbox{{\boldmath ${y}$}$_{i-1}$}|f{^{\prime}},%
\mbox{{\boldmath
${X}$}$_{i-1}$})dp_{\theta_0}(f{^{\prime}})}  \notag \\
=&{\int_{{\mathcal{F}}} p_{\phi_0}(\mbox{{\boldmath ${y}$}$_{n}$}|f, %
\mbox{{\boldmath ${X}$}$_{n}$})dp_{\theta_0}(f)}=p_{\phi_0,\theta_0}(%
\mbox{{\boldmath ${y}$}$_{n}$} |\mbox{{\boldmath ${X}$}$_{n}$}),  \notag
\end{align}
which shows that the equation (\ref{bspred}) holds.$\sharp$

\vskip 10pt \noindent \textbf{Lemma 3} \textit{Suppose $%
\mbox{{\boldmath
${y}$}$_{n}$}=\{y_1,...,y_n\}$ are generated from the eTPR model (\ref{assum1}%
) with the mean function $h(\mbox{\boldmath ${x}$})=0$, and covariance
kernel function $k$ is bounded and continuous in parameter $%
\mbox{\boldmath
${\theta}$} $. It also assumes that the estimate $\hat{%
\mbox{\boldmath
${\beta}$}}$ almost surely converges to $\mbox{\boldmath ${\beta}$}$ as $%
n\rightarrow \infty$. Then for a positive constant $c$, and any $\varepsilon>0$%
, when $n$ is large enough, we have
\begin{align}
&\frac{1}{n}(-\log p_{\phi_0,\hat\theta}(\mbox{{\boldmath ${y}$}$_{n}$}|%
\mbox{{\boldmath ${X}$}$_{n}$})+\log p_{\phi_0}( \mbox{{\boldmath
${y}$}$_{n}$}|f_0,\mbox{{\boldmath ${X}$}$_{n}$}))  \notag \\
\leq& \frac{1}{n}\left\{\frac{1}{2}\log|\mbox{{\boldmath ${I}$}$_{n}$}%
+\phi_0^{-1} \mbox{{\boldmath ${K}$}$_{n}$}|+ \frac{s^2+2(\nu-1)}{2(n+2\nu-2)%
}(||f_0||^2_k+c)+c\right\}+ \varepsilon,  \notag
\end{align}
where $\mbox{{\boldmath ${K}$}$_{n}$}=(k(x_i,x_j))_{n\times n}$, $s^2=(%
\mbox{{\boldmath ${y}$}$_{n}$}-f_0(\mbox{{\boldmath ${X}$}$_{n}$}))^T(%
\mbox{{\boldmath ${y}$}$_{n}$}-f_0(\mbox{{\boldmath ${X}$}$_{n}$}))/\phi_0$,
$\mbox{{\boldmath ${I}$}$_{n}$}$ is the $n\times n$ identity matrix, and $%
||f_0||_k$ is the reproducing kernel Hilbert space norm of $f_0$ associated
with kernel function $k(\cdot,\cdot;\mbox{\boldmath ${\theta}$})$. }

\vskip 10pt \textbf{Proof}: From Proposition 1, it follows that there exists a
variable $r\sim IG(\nu,(\nu-1))$ with density function $g(r)$, conditional
on $r$ we have
\begin{align}
\left(
\begin{array}{c}
f \\
\epsilon
\end{array}
\right)\Big|r\sim GP\left(\left(
\begin{array}{c}
0 \\
0
\end{array}
\right),\left(
\begin{array}{cc}
r k & 0 \\
0 & r \tilde{k}
\end{array}
\right)\right),  \notag
\end{align}
where $GP(h,k)$ stands for Gaussian process with mean function $h$ and
covariance function $k$. Then conditional on $r$, the extended t-process
regression model (\ref{assumed}) becomes Gaussian process regression model
\begin{eqnarray}  \label{rGPR}
y(\mbox{\boldmath ${x}$})=\tilde{f}(\mbox{\boldmath ${x}$})+\tilde{\epsilon}(%
\mbox{\boldmath ${x}$}),
\end{eqnarray}
where $\tilde{f}=f|r\sim GP(0, rk(\cdot,\cdot;\theta))$, $\tilde{\epsilon}%
|r\sim GP(0,r\tilde{k}(\cdot,\cdot;\phi_0))$ , and $\tilde{f} $ and error
term $\tilde{\epsilon}$ are independent. Denoted $\tilde{p}$ by probability
density computation conditional on $r$. Based on the model (\ref{rGPR}), let
\begin{align}
&p_{G}(\mbox{{\boldmath ${y}$}$_{n}$}|r,\mbox{{\boldmath ${X}$}$_{n}$}%
)=\int_{\mathcal{F}}\tilde{p}_{\phi_0}(\mbox{{\boldmath ${y}$}$_{n}$}|\tilde{%
f},\mbox{{\boldmath ${X}$}$_{n}$}) d\tilde{p}_n(\tilde{f}),  \notag \\
&p_{0}(\mbox{{\boldmath ${y}$}$_{n}$}|r,\mbox{{\boldmath ${X}$}$_{n}$})=%
\tilde{p}_{\phi_0}(\mbox{{\boldmath
${y}$}$_{n}$}|f_0,\mbox{{\boldmath ${X}$}$_{n}$}),  \notag
\end{align}
where $\tilde{p}_n$ is the induced measure from Gaussian process $%
GP(0,rk(\cdot,\cdot;\hat{\mbox{\boldmath ${\theta}$}}))$.

We know that random effect $r$ is independent of covariates $%
\mbox{{\boldmath ${X}$}$_{n}$}$. Then it easily shows that
\begin{align}
&p_{\phi_0,\hat{\theta}}(\mbox{{\boldmath ${y}$}$_{n}$}|%
\mbox{{\boldmath
${X}$}$_{n}$})=\int p_{G}(\mbox{{\boldmath
${y}$}$_{n}$}|r,\mbox{{\boldmath ${X}$}$_{n}$}) g(r) dr,  \label{tpr1} \\
&p_{\phi_0}(\mbox{{\boldmath
${y}$}$_{n}$}|f_0,\mbox{{\boldmath ${X}$}$_{n}$})=\int p_{0}(%
\mbox{{\boldmath
${y}$}$_{n}$}|r,\mbox{{\boldmath ${X}$}$_{n}$}) g(r) dr.  \label{tpr0}
\end{align}

Suppose that for any given $r$, we have
\begin{align}  \label{GPcons}
&-\log p_{G}(\mbox{{\boldmath ${y}$}$_{n}$}|r,\mbox{{\boldmath ${X}$}$_{n}$}%
)+\log p_{0}(\mbox{{\boldmath ${y}$}$_{n}$}|r,\mbox{{\boldmath ${X}$}$_{n}$})
\notag \\
\leq& \frac{1}{2}\log|\mbox{{\boldmath ${I}$}$_{n}$}+\phi_0^{-1} %
\mbox{{\boldmath ${K}$}$_{n}$}|+ \frac{r}{2}(||f_0||^2_k+c)+c+n\varepsilon,
\end{align}
which indicates
\begin{align}  \label{loggp}
&-\log\int p_{G}(\mbox{{\boldmath ${y}$}$_{n}$}|r,%
\mbox{{\boldmath
${X}$}$_{n}$}) g(r) dr \leq \frac{1}{2}\log|\mbox{{\boldmath ${I}$}$_{n}$}%
+\phi_0^{-1} \mbox{{\boldmath ${K}$}$_{n}$}|+c +n\varepsilon  \notag \\
&\hskip 2cm -\log \int p_{0}(\mbox{{\boldmath ${y}$}$_{n}$}|r,%
\mbox{{\boldmath ${X}$}$_{n}$}) \exp\{-( \frac{r}{2}(||f_0||^2_k+c))\} g(r)
dr.
\end{align}
By simple computation, it follows that
\begin{align}  \label{gtilde}
&\int p_{0}(\mbox{{\boldmath ${y}$}$_{n}$}|r,\mbox{{\boldmath ${X}$}$_{n}$})
\exp\{-( \frac{r}{2}(||f_0||^2_k+c))\} g(r) dr  \notag \\
=&\int p_{0}(\mbox{{\boldmath ${y}$}$_{n}$}|r,\mbox{{\boldmath ${X}$}$_{n}$}%
) g(r) dr \int \exp\{-( \frac{r}{2}(||f_0||^2_k+c))\} \tilde{g}(r) dr,
\end{align}
where $\tilde{g}(r)$ is the density function of $IG(\nu+n/2,(%
\nu-1)+s^2/2)$. From (\ref{tpr1}), (\ref{tpr0}), (\ref{loggp}) and (\ref
{gtilde}), we have
\begin{align}
&-\log p_{\phi_0,\hat{\theta}}(\mbox{{\boldmath
${y}$}$_{n}$}|\mbox{{\boldmath ${X}$}$_{n}$})+\log p_{\phi_0}( %
\mbox{{\boldmath ${y}$}$_{n}$}|f_0,\mbox{{\boldmath ${X}$}$_{n}$})  \notag \\
\leq& \frac{1}{2}\log|\mbox{{\boldmath ${I}$}$_{n}$}+\phi_0^{-1} %
\mbox{{\boldmath ${K}$}$_{n}$}|+c-\log \int \exp\{-( \frac{r}{2}%
(||f_0||^2_k+c))\} \tilde{g}(r) dr  \notag \\
\leq &\frac{1}{2}\log|\mbox{{\boldmath ${I}$}$_{n}$}+\phi_0^{-1} %
\mbox{{\boldmath ${K}$}$_{n}$}|+c+\frac{||f_0||^2_k+c}{2} \int r \tilde{g}%
(r) dr  \notag \\
=&\frac{1}{2}\log|\mbox{{\boldmath ${I}$}$_{n}$}+\phi_0^{-1}
\mbox{{\boldmath
${K}$}$_{n}$}|+ \frac{s^2+2(\nu-1)}{2(n+2\nu-2)}(||f_0||^2_k+c)+c+n%
\varepsilon,  \notag
\end{align}
which shows that Lemma 3 holds.

Now let us prove the inequality (\ref{GPcons}). Since the proof of (\ref
{GPcons}) is similar to those of Theorem 1 in Seeger \textit{et al}. (2008) and Lemma
1 in Wang and Shi (2014), here we summarily present the procedure of the
proof, details please see in Seeger \textit{et al}. (2008) and Wang and Shi (2014).
Let $\mathcal{H}$ be the reproducing kernel Hilbert space (RKHS) associated
with covariance function $k(\cdot,\cdot;\mbox{\boldmath ${\theta}$})$, and $%
\mathcal{H}_n= \{\tilde{f}(\cdot): \tilde{f}(\cdot)=\sum_{i=1}^n\alpha_i k(%
\mbox{\boldmath ${x}$},\mbox{{\boldmath ${x}$}$_{i}$};\mbox{\boldmath
${\theta}$})$, for any $\alpha_i\in R\}$. From the Representer Theorem (see
Lemma 2 in Seeger \textit{et al}., 2008), it is sufficient to prove (\ref{GPcons})
for the true underlying function $\tilde{f}_0=f_0|r\in \mathcal{H}_n$. Then
for given $r$, ${f}_0 $ can be written as
\begin{align}
&{f}_0(\cdot)=r\sum_{i=1}^n\alpha_i k(\mbox{\boldmath ${x}$},%
\mbox{{\boldmath ${x}$}$_{i}$};\mbox{\boldmath ${\theta}$})\doteq r K(\cdot)%
\mbox{\boldmath ${\alpha}$},  \notag
\end{align}
where $K(\cdot)=(k(\mbox{\boldmath ${x}$},\mbox{{\boldmath ${x}$}$_{1}$};%
\mbox{\boldmath ${\theta}$}),...,k(\mbox{\boldmath ${x}$},%
\mbox{{\boldmath
${x}$}$_{n}$};\mbox{\boldmath ${\theta}$}))$ and $\mbox{\boldmath ${\alpha}$}%
=(\alpha_1,...,\alpha_n)^T$.

By Fenchel-Legendre duality relationship, we have
\begin{align}  \label{FLD}
-\log p_{G}(\mbox{{\boldmath ${y}$}$_{n}$}|r,\mbox{{\boldmath ${X}$}$_{n}$})
\leq E_{Q}(-\log \tilde{p}(\mbox{{\boldmath ${y}$}$_{n}$}|\tilde{f}))+D[Q,P],
\end{align}
where $P$ is a measure induced by $GP(0,rk(\cdot,\cdot;\hat{%
\mbox{\boldmath
${\theta}$}}_{n}))$, and $Q$ is the posterior distribution of $\tilde f$
from a GP model with prior $GP(0, rk(\cdot,\cdot;\mbox{\boldmath ${\theta}$}%
))$ and Gaussian likelihood term $\prod_{i=1}^n N(\hat y_i|\tilde f(%
\mbox{{\boldmath ${x}$}$_{i}$}),r\phi_0)$, where $\hat{\mathbf{y}}=(\hat
y_1,...,\hat y_n)^T=r( \mbox{{\boldmath ${K}$}$_{n}$}+\phi_0%
\mbox{{\boldmath
${I}$}$_{n}$})\mbox{\boldmath ${\alpha}$}$ and $%
\mbox{{\boldmath
${K}$}$_{n}$}=(k(\mbox{{\boldmath ${x}$}$_{i}$},%
\mbox{{\boldmath
${x}$}$_{j}$};\mbox{\boldmath ${\theta}$})$. Then we have $E_{Q}(\tilde
f)=f_0$. Let $\mbox{\boldmath ${B}$}=\mbox{{\boldmath ${I}$}$_{n}$}%
+\phi_0^{-1} \mbox{{\boldmath ${K}$}$_{n}$}$, then we have
\begin{align}
&D[Q,P]=\frac{1}{2}\left\{-\log|\hat{\mbox{\boldmath ${K}$}}_{n}^{-1} %
\mbox{{\boldmath ${K}$}$_{n}$}|+\log|\mbox{\boldmath ${B}$}|+tr(\hat{%
\mbox{\boldmath ${K}$}}_{n}^{-1} \mbox{{\boldmath ${K}$}$_{n}$}%
\mbox{{\boldmath ${B}$}$^{-1}$})\right.  \notag \\
&\hskip 2cm \left.+r||f_0||^2_k+r\mbox{\boldmath ${\alpha }$} %
\mbox{{\boldmath ${K}$}$_{n}$}(\hat{\mbox{\boldmath ${K}$}}_{n}^{-1} %
\mbox{{\boldmath ${K}$}$_{n}$}-\mbox{{\boldmath ${I}$}$_{n}$})%
\mbox{\boldmath ${\alpha}$}-n\right\},  \label{KBdist} \\
&E_{Q}(-\log \tilde{p}(\mbox{{\boldmath ${y}$}$_{n}$}|\tilde{f}))\leq -\log
\tilde{p}(\mbox{{\boldmath ${y}$}$_{n}$}|f_0)+\frac{1}{2} \phi_0^{-1}tr( %
\mbox{{\boldmath ${K}$}$_{n}$}\mbox{{\boldmath ${B}$}$^{-1}$})  \notag \\
&\hskip 3.2cm =-\log p_{0}(\mbox{{\boldmath ${y}$}$_{n}$}|r,%
\mbox{{\boldmath
${X}$}$_{n}$})+\frac{1}{2} \phi_0^{-1}tr( \mbox{{\boldmath ${K}$}$_{n}$}%
\mbox{{\boldmath ${B}$}$^{-1}$}),  \label{EQP}
\end{align}
where $\hat{\mbox{\boldmath ${K}$}}_{n}=(k(\mbox{{\boldmath ${x}$}$_{i}$},%
\mbox{{\boldmath ${x}$}$_{j}$};\hat{\mbox{\boldmath ${\theta}$}}))$.

Hence, it follows from (\ref{FLD}), (\ref{KBdist}) and (\ref{EQP}) that
\begin{align}  \label{GPcons-1}
&-\log p_{G}(\mbox{{\boldmath ${y}$}$_{n}$}|r,\mbox{{\boldmath ${X}$}$_{n}$}%
)+\log p_{0}(\mbox{{\boldmath ${y}$}$_{n}$}|r,\mbox{{\boldmath ${X}$}$_{n}$})
\notag \\
\leq& \frac{1}{2}\left\{-\log|\hat{\mbox{\boldmath ${K}$}}_{n}^{-1} %
\mbox{{\boldmath ${K}$}$_{n}$}|+\log|\mbox{\boldmath ${B}$}|+tr((\hat{%
\mbox{\boldmath ${K}$}}_{n}^{-1} \mbox{{\boldmath ${K}$}$_{n}$}+\phi_0^{-1} %
\mbox{{\boldmath ${K}$}$_{n}$})\mbox{{\boldmath ${B}$}$^{-1}$})
+r||f_0||^2_k\right.  \notag \\
&\left.+r\mbox{\boldmath ${\alpha }$} \mbox{{\boldmath ${K}$}$_{n}$}(\hat{%
\mbox{\boldmath ${K}$}}_{n}^{-1} \mbox{{\boldmath ${K}$}$_{n}$}-%
\mbox{{\boldmath ${I}$}$_{n}$})\mbox{\boldmath ${\alpha}$}-n\right\}.
\end{align}
Since the covariance function is bounded and continuous in $%
\mbox{\boldmath
${\theta}$}$ and $\hat{\mbox{\boldmath ${\theta}$}}\rightarrow %
\mbox{\boldmath ${\theta}$}$, we have $\hat{\mbox{\boldmath ${K}$}}_{n}^{-1} %
\mbox{{\boldmath ${K}$}$_{n}$} -\mbox{{\boldmath ${I}$}$_{n}$}\rightarrow 0$
as $n\rightarrow \infty$. Hence, there exist positive constants $c$ and $%
\varepsilon$ such that for $n$ large enough
\begin{align}  \label{eqns}
&-\log|\hat{\mbox{\boldmath ${K}$}}_{n}^{-1} \mbox{{\boldmath ${K}$}$_{n}$}%
|<c,~~\mbox{\boldmath ${\alpha }$} \mbox{{\boldmath ${K}$}$_{n}$}(\hat{%
\mbox{\boldmath ${K}$}}_{n}^{-1} \mbox{{\boldmath ${K}$}$_{n}$}-%
\mbox{{\boldmath ${I}$}$_{n}$})\mbox{\boldmath ${\alpha}$}<c,  \notag \\
&tr(\hat{\mbox{\boldmath ${K}$}}_{n}^{-1} \mbox{{\boldmath ${K}$}$_{n}$} %
\mbox{{\boldmath ${B}$}$^{-1}$})<tr((\mbox{{\boldmath ${I}$}$_{n}$}%
+\varepsilon \mbox{{\boldmath ${K}$}$_{n}$})\mbox{{\boldmath ${B}$}$^{-1}$}).
\end{align}
Plugging (\ref{eqns}) in (\ref{GPcons-1}), we have the inequality (\ref
{GPcons}). $\sharp$

\vskip 10pt

For proof of Proposition 3, we need condition \newline
(A) $||f_{0}||_{k}$ is bounded and $E_{\mbox{{\boldmath
${X}$}$_{n}$}}(\log|\mbox{{\boldmath ${I}$}$_{n}$}+\phi_0^{-1}%
\mbox{{\boldmath
${K}$}$_{n}$}|)=o(n)$.

\noindent\textbf{Proof of Proposition 3}: It
easily shows that $s^2=(\mbox{{\boldmath ${y}$}$_{n}$}-f_0(\mbox{{\boldmath
${X}$}$_{n}$}))^T(\mbox{{\boldmath ${y}$}$_{n}$}-f_0(\mbox{{\boldmath
${X}$}$_{n}$}))/\phi_0=O(n) $. Under conditions in Lemma 3, and condition
(A), it follows from Lemma 3 that
\begin{equation}
\frac{1}{n}E_{\mbox{{\boldmath ${X}$}$_{n}$}}(D[p_{\phi _{0}}(%
\mbox{{\boldmath
${y}$}$_{n}$}|f_{0},\mbox{{\boldmath ${X}$}$_{n}$}),p_{\phi _{0},\hat{\theta}%
}(\mbox{{\boldmath ${y}$}$_{n}$}|\mbox{{\boldmath ${X}$}$_{n}$}%
)])\longrightarrow 0,{\mbox{as}}~~n\rightarrow \infty .  \notag
\end{equation}
That proves Proposition 3.$\sharp$

\section*{\textbf{References}}

\begin{enumerate}
\item  Archambeau, C. and Bach, F. (2010), Multiple Gaussian Process Models.
\textit{Advances in Neural Information Processing Systems}.

\item  Arellano-Valle, R. B. and Bolfarine, H. (1995). On some
characterization of the t-distribution. \textit{Statistics} \& \textit{%
Probability Letters}, 25, 79 - 85.



\item  Cleveland, W.S., Devlin, S.J. (1988), Locally-Weighted Regression: An
Approach to Regression Analysis by Local Fitting. \textit{Journal of the
American Statistical Association} 83: 596-610.

\item  Dutta, S. and Mondal, D. (2015), An h-likelihood method for spatial
mixed linear models based on intrinsic auto-regressions. \textit{J. R.
Statist. Soc. B} 77: 699-726.


\item  Hall, P., M\"{u}ller, H.-G., and Yao, F. (2008), Modelling Sparse
Generalized Longitudinal Observations with Latent Gaussian Processes,\textit{%
Journal of Royal Statistical Society, Ser. B}, 70, 703-723.

\item  Hampel, F.R., Ronchetti, E.M., Rousseeuw, P.J. and Stahel, W.A.
(1986), Robust Statistics: The Approach Based on Influence Functions, Wiley.



\item  Lange, K.L., Little, R. J.A. and Taylor J. M.G. (1989), Robust
statistical modelling using the t distribution, \textit{Journal of the
American Statistical Association}, 84, 881-896.

\item  Lee, Y. and Kim, G. (2015). H-likelihood predictive intervals for
unobservables, \textit{International Statistical Review}, DOI:
10.1111/insr.12115.

\item  Lee, Y. and Nelder, J.A. (1996). Hierarchical Generalized Linear
Models. \textit{Journal of the Royal Statistical Society B}, 58, 619-678.

\item  Lee, Y. and Nelder, J.A. (2006). Double hierarchical generalized
linear models (with discussion). \textit{Journal of the Royal Statistical
Society: C (Applied Statistics)}, 55, 139-185.

\item  Lee, Y., Nelder, J.A. and Pawitan, Y. (2006). Generalized Linear
Models with Random Effects, Unified Analysis via H-likelihood. Chapman \&
Hall/CRC.

\item  Ma, R. and Jorgensen, B. (2007). Nested generalized linear mixed
models: an orthodox best linear unbiased predictor approach. \textit{Journal
of the Royal Statistical Society B}, 69, 625-641.



\item  Rasmussen, C. E. and Williams, C. K. I. (2006), Gaussian Processes
for Machine Learning. Cambridge, Massachusetts: The MIT Press.

\item  Robinson, G.K. (1991). That BLUP is a good thing: the estimation of
random effects (with discussion). \textit{Statistical Science} 6, 15-51.



\item  Seeger M. W., Kakade S. M. and Foster D. P. (2008), Information
Consistency of Nonparametric Gaussian Process Methods, IEEE Transactions on
Information Theory, 54, 2376-2382.

\item  Shah A., Wilson A.G. and Ghahramani Z. (2014), Student-t processes as
alternatives to Gaussian processes. Proceedings of the 17th International
Conference on Artificial Intelligence and Statistics (AISTATS), 877-885.

\item  Shi, J. Q. and Choi, T. (2011), Gaussian Process Regression Analysis
for Functional Data, London: Chapman and Hall/CRC.




\item  Wang, B. and Shi, J.Q. (2014), Generalized Gaussian process
regression model for non-Gaussian functional data. \textit{Journal of the
American Statistical Association}, 109, 1123-1133.

\item  Wauthier, F. L. and Jordan, M. I. (2010). Heavy-tailed process priors
for selective shrinkage. In Advances in Neural Information Processing
Systems, 2406-2414.

\item  Xu, P, Lee, Y. and Shi, J. Q. (2015) Automatic Detection of
Significant Areas for Functional Data with Directional Error Control.
arXiv:1504.08164.

\item  Xu, Z., Yan, F. and Qi, Y. (2011), Sparse Matrix-Variate t Process
Blockmodel. Proceedings of the 25th AAAI Conference on Artificial
Intelligence, 543-548.

\item  Yu S., Tresp V. and Yu K. (2007), Robust multi-tast learning with
t-process. Proceedings of the 24th International Conference on Machine
Learning, 1103-1110.

\item  Zellener, A. (1976), Bayesian and non-Bayesian analysis of the
regression model with multivariate student-t error terms, \textit{Journal of
the American Statistical Association}, 71, 400-405.

\item  Zhang, Y. and Yeung, D.Y. (2010), Multi-task learning using
generalized $t$ process. Proceedings of the 13th International Conference on
Artificial Intelligence and Statistics (AISTATS), 964-971.

\item  Zhou, X. and Stephens, M. (2012), Genome-wide efficient mixed-model
analysis for association studies. \textit{Nature Genetics} 44: 821-824.
\end{enumerate}

\end{document}